\documentclass[12pt]{article}

\usepackage{amsmath}
\usepackage{graphicx}
\usepackage{enumerate}
\usepackage[numbers, sort&compress]{natbib}
\usepackage{url} % not crucial - just used below for the URL 
\usepackage{rotating}
\usepackage{array}
\usepackage{longtable}
\usepackage{booktabs}
\usepackage{graphicx}
\usepackage{caption}
\usepackage{calc}

\PassOptionsToPackage{hidelinks}{hyperref}
\usepackage{hyperref}

\usepackage{pdflscape}
\newcommand{\blandscape}{\begin{landscape}}
	\newcommand{\elandscape}{\end{landscape}}

\DeclareMathOperator\dd{d}
\newcommand{\ind}{{\perp\!\!\!\perp}}

\providecommand{ }{%
	\setlength{\itemsep}{0pt}\setlength{\parskip}{0pt}}

\title{Evaluating treatment effects on longitudinal outcomes with attrition due to death: Methods for a two-dimensional while-alive estimand}

\author{Dries Reynders$^1$, Doranne Thomassen$^2$, Satrajit Roychoudhury$^3$,\\ Cecilie Delphin Amdal$^{4,5}$,Jammbe Z. Musoro$^6$, Willi Sauerbrei$^7$,\\ Saskia le Cessie$^{1,2,8}$ , Els Goetghebeur$^1$;\\ 
	on behalf of the SISAQOL-IMI Work Package 3  members  listed in Appendix\\
}

\date{ \small
	$^1$ Department of Applied Mathematics, Computer Science and Statistics, Ghent
	University, Ghent, Belgium.\\
	$^2$ Department of Biomedical Data Sciences, Leiden University Medical Center,
	P.O. box 9600, Postzone S-05-S, Leiden 2300 RC, The Netherlands.\\
	$^3$ Pfizer Inc, New York, NY, USA. \\
	$^4$ Research Support Services, Oslo University Hospital, Oslo, Norway. \\
	$^5$ Department of Oncology, Oslo University Hospital, Oslo, Norway. \\
	$^6$ European Organisation for Research and Treatment of Cancer (EORTC) Headquarters, Brussels, Belgium. \\
	$^7$ Institute of Medical Biometry and Statistics, Faculty of Medicine and Medical Center, University of Freiburg, Freiburg, Germany. \\
	$^8$ Department of Clinical Epidemiology, Leiden
	University Medical Center, Leiden, The Netherlands}

\begin{document}

\maketitle

\begin{abstract}
\textbf{Background:} In clinical studies, substantial mortality complicates the analysis of longitudinal outcomes since they do not exist after death. This poses challenges for defining, estimating and interpreting causal effects. Common approaches incorporate survival and longitudinal outcomes into a composite endpoint or invoke additional assumptions to accommodate truncation by death. Yet, many questions of clinical and policy relevance require explicit recognition of survival as a distinct outcome dimension rather than viewing death as a missing-data mechanism.

\textbf{Methods:} We propose a two-dimensional while-alive estimand that summarizes the joint distribution of survival and longitudinal outcomes among survivors at scheduled times. This estimand explicitly acknowledges survival as an integral component of the outcome and provides a meaningful target for causal inference. We compare two estimation approaches applicable in observational studies, assuming no unmeasured confounding for treatment and censoring. Regression-standardization models both survival and longitudinal outcomes and standardizes them to a baseline target population, weighting longitudinal outcome expectations by the estimated survival distribution. 'Inverse probability of treatment and censoring weighting' estimates weighted mean outcomes among observed survivors using separate weights to adjust for baseline confounding and time-dependent censoring. Their finite-sample performance is evaluated in a simulation study. The methods are further illustrated in an oncology case study evaluating quality of life (QoL). A negative-control analysis examines the underlying assumptions.

\textbf{Results:} Simulations showed negligible bias and close-to-nominal confidence interval coverage under correct model specification. In the case study, the treatment groups exhibited similar survival, while the experimental treatment yielded substantially improved QoL among survivors. The negative control analysis found no signs of bias. Together, these findings support a favourable assessment of the experimental intervention. Had estimated effects on survival and QoL while-alive favoured different treatments, the treatment preference would follow a context-specific evaluation of the relative importance of these results.

\textbf{Conclusion:} The proposed two-dimensional while-alive estimand addresses questions important to patients, caregivers and clinicians. Treating survival as an outcome component rather than a source of selection bias facilitates transparent interpretation and provides an actionable summary of causal treatment effects. The proposed estimation procedures exhibit favourable statistical performance and can be implemented using standard software.
\end{abstract}

\section{Introduction}

In many clinical studies, the outcome of interest is measured repeatedly over time. The analysis becomes particularly complex when terminal events such as death occur during the study period. Consider, for example, the impact of an oncology treatment on  Quality of Life (QoL), the evaluation of a weight loss drug in late stage diabetes, or physical and mental functioning across care home residents. Since longitudinal outcomes no longer exist after death, mortality requires explicit consideration when defining the estimand and its estimator. Arguably, key questions for patients and caregivers are then: '\textit{When starting that treatment, what are the chances to remain alive?}' and '\textit{Having started that treatment, what can be expected regarding quality of life while still alive?}'. This set of questions acknowledges that survival and longitudinal responses are intrinsically intertwined and jointly define a two-dimensional outcome process.

The ICH-E9(R1) \cite{europeanmedicinesagency2020} estimand framework outlines several strategies to address death when estimating treatment effects in clinical trials: hypothetical, composite endpoint, principal stratum and while-alive. Comparing outcomes only between those still alive, is often considered subject to selection bias through differential survival and hence without causal interpretation \cite{egleston2007, vanderweele2011, shardell2015,   wang2017, colantuoni2018}. The first three strategies aim to circumvent this presumed problem by modifying either the outcome or the target population. Below, we label estimands following the strategy for handling death. Hypothetical estimands then target an effect in a counterfactual world where death is avoided. This requires (implicitly) imputing values after death and evaluating the resulting augmented outcome across the entire population that entered the study at baseline. Composite estimands integrate death with the longitudinal outcome into one scale, for example by assigning a QoL score zero after death or by targeting 'time to QoL deterioration or death'. By doing so, they no longer distinguish the unique state of death from living with a low quality of life. Principal stratum estimands instead target an effect on the longitudinal outcome in a generally unidentifiable (cross worlds) sub-population: among subjects that would remain alive under either treatment at considered time points. Nevertheless, standard analyses based on (Generalized) Linear Mixed Models typically report marginal effects corresponding to a hypothetical estimand \cite{rouanet2019, kurland2005, kurland2009}. Meanwhile, the methodological literature on truncation by death has predominantly considered composite endpoint approaches \cite{wang2017, xiang2023} or principal stratification estimands defined within survivor subpopulations \cite{luo2023,shardell2015, tchetgentchetgen2012, egleston2007}.

\begin{figure}[htbp!]
	\begin{center}
		\includegraphics[width = 0.45\textwidth]{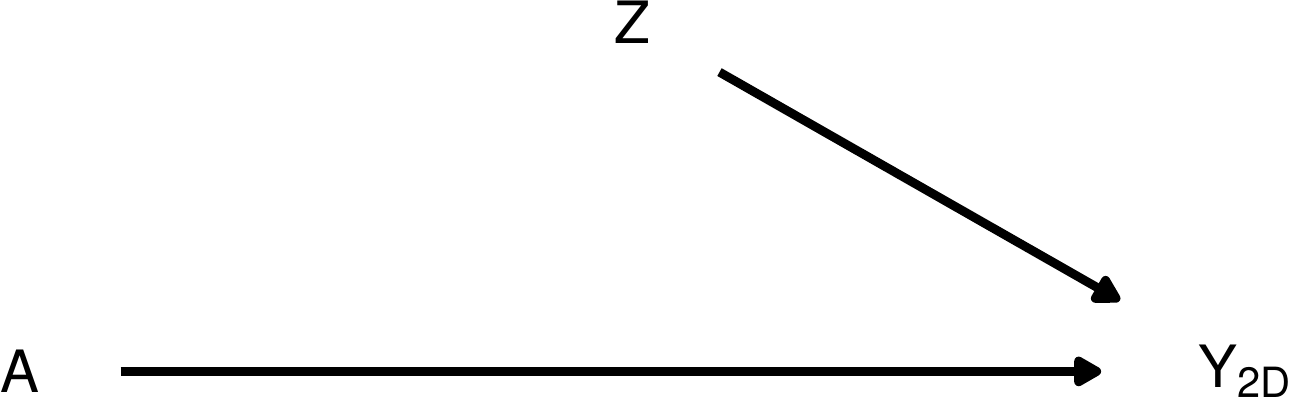}    
		\caption{When the two-dimensional outcome is assessed as a whole, the selection through survival is an integral part of the outcome and not a source of bias \label{2dimDAG}}
	\end{center}
\end{figure}

'While-alive' strategies can target a range of summary measures (see e.g. \cite{pham2025, kahan2024}). Our proposed two-dimensional while-alive estimand considers the two-dimensional outcome $Y_{2D}= (T, \mathbf{\overline{Y}}(T^-))$ consisting of survival time $T$ and $\mathbf{\overline{Y}}(T^-)$, the history of longitudinal outcomes while alive at scheduled measurement-times $t_j, j \in \{0,1,\ldots\}$. It presents survival probabilities over relevant follow-up times before summarizing the longitudinal outcome among the living at these scheduled times. Figure \ref{2dimDAG} presents a simple causal Directed Acyclic Graph of a randomized experiment with treatment assignment $A$, prognostic factors $Z$ and outcome $Y_{2D}$. It visualizes how differences in $Y_{2D}$, the joint distribution of survival and the longitudinal outcome, are seen as causal effects. Rather than selection bias, differential survival is a consequence of treatment choice, as is the health status among patients alive. Hence, contrasting the proposed summary of $Y_{2D}$, possibly stratified on baseline covariates, presents a causal effect on what patients may experience following treatment, and can inform  the initial treatment decision. Where the survival component is linked to '\textit{chances to remain alive}', the expected outcome among survivors represents '\textit{what can be expected if still alive}'.

Motivated by the relevance of this causal two-dimensional while-alive estimand, we implement two intuitive methods for estimation from observational data. We thereby assume no unmeasured baseline confounding and independent censoring conditional on possibly time-varying covariates.

The first method applies Regression-Standardization \cite{robins1986,hernan2020}, first regressing each outcome dimension $T$ and $Y$  on baseline covariates and then standardizing to a target population of interest. The second method applies inverse probability weighting \cite{robins2000,hernan2020} twice: 1) to obtain baseline covariate balance between treatment groups and 2) to account for censoring. Both approaches can be seen as extensions of 'regression conditional on being alive' or 'partly conditional regression'\cite{kurland2005,dufouil2004,wen2018a}, where both survival and the longitudinal outcome are standardized towards a target population.

We start by introducing notation and formally define a causal two-dimensional while-alive estimand for a target population of interest in Section 2. In Section 3 we develop two corresponding estimators for an observational setting. The main ideas are developed by targeting a single time point evaluation first and then extended to a fully longitudinal analysis. In Section 4 a small simulation study examines the methods' performance in finite samples. The methods are then applied in a case study in Section 5, where the effect of an anticancer drug on QoL is evaluated by comparing single arm trial data with an external control.

\section{Preliminaries} \label{sec-notprel}
\subsection{Notation}

Consider two independent samples $S_1$ and $S_0$ from populations $\mathcal{P}_1$ and $\mathcal{P}_0$, of patients taking the treatment under study and standard of care, respectively. We consider treatment assignment $A$ (1 for experimental treatment, 0 for control), time to death (censoring) $T$ ($C$) with $X=\mathrm{min}(T,C)$, and failure indicator $\delta$. The continuous outcome $Y(t)$ only exists if $t < T$ and is measured among survivors at planned times $t_j$ with $j\in \{0,1,\ldots, J\}$. $\overline {\mathbf {Y}}(t)$, is the vector of repeated outcomes $Y(t_j), t_j \leq t$. $f_{\mathcal{P}}$ and $E_{\mathcal{P}}$ refer respectively to a density and expectation over a population $\mathcal{P}$. Sampled individuals are indexed $i$, with $i \in \{1,2,\ldots, m_0\}$ for those in $S_0$ and $i \in \{m_0 + 1,m_0 +2,\ldots, m_0+m_1\}$ for those in $S_1$. With a slight abuse of notation, we will use the shorthand $i\in S_{0/1}$ to indicate the corresponding sets of indices. An overview of the notation is provided in table \ref{tbl-notation}.

\begin{table}[ht]
	\caption{Overview of Notation}
	\centering
	\begin{tabular}{@{}p{2.3cm}p{13cm}@{}}
		\toprule
		\textbf{Variable} & \textbf{Description} \\
		\midrule
		$A$ &  Treatment-indicator 0 (control) or 1 (experimental treatment) prescribed or otherwise assigned at time $t=0$ \\
		$S_a$ & Sample of patients with $A=a$\\
		$T$ & Time to death, measured from a well-defined time origin $t=0$ (e.g. disease onset or study enrolment) \\
		$Y(t)$ &  Continuous outcome of interest at time $t$ where measurements are scheduled to be taken at pre-planned time points \(t_j, j \in \{0,...J\}\) prior to $T$. At and after $T$, $Y$ is no longer defined. \\
		$\overline{\mathbf{Y}}(t)$ & Vector of repeated measures $Y(t_j)$  for times $t_j \leq t$ \\
		$C$ & Time to censoring, measured on the same time scale and with the same time origin as $T$ \\
		$X$ & $= \text{min}(T,C)$  the observed follow-up time for survival \\
		$\delta$ & Failure indicator 1 (death observed) or 0 (censored) \\
		$\mathbf{Z}$ & A vector of baseline covariates\\
		$t^*$ & Fixed time point of interest \\
		${\cal{P}^*}$ & Target population, the population for which we want to estimate the average treatment effect \\
		$S^*$ & A representative sample of $\cal{P}^*$ \\
		${\cal{P}}_a$ & Population for which $S_a$ is a representative sample\\
		$E_{\cal{P}}$, $f_{\cal{P}}$ & Population $\cal{P}$-specific expected value and density, respectively \\
		$(T^a, \overline{\mathbf{Y}}^a(T^{a-}))$ & Two-dimensional potential outcomes \\
		\bottomrule
	\end{tabular}
	\label{tbl-notation}
\end{table}

To formalize causal effects we introduce potential outcomes \cite{goetghebeur2020} under possible treatments $a=0,1$ (assuming consistency and no interference as defined in Section \ref{sec-estimation}):$(T_i^a, \overline {\mathbf{Y}}_i^a(T_i^{a-}))$,  individual $i$'s set of potential outcomes under either treatment assignment \(a = 0, 1\) of which only $a=a_i$ is observed.

\subsection{The Two-Dimensional While-Alive Estimand} \label{sec-estimand}

We define five components of our two-dimensional causal while-alive estimand as follows: the target population ${\cal{P}}^*$ (e.g. corresponding to the treated population ${\cal{P}}_1$). Initiating the experimental treatment ($A=1$) vs. standard of care ($A=0$) is considered as point-treatment, with survival time $T$ and $Y$ at a specific time $t^*$ among survivors at that time, as outcomes of interest. All intercurrent events apart from death are handled with a treatment-policy strategy. We are hence interested in the outcome regardless of whether patients stopped the treatment or had disease progression, but the longitudinal outcome ceases to exist after death. The population level summaries are (the difference in) survival curves following the respective assigned treatments presenting the first dimension and (difference in) expected value of $Y(t^*)$ among those still alive at $t^*$ in ${\cal{P}}^* $ presenting the second dimension. This can be evaluated for different values of $t^*$:

\begin{align}\label{eq-estimand2}
	&P_{\cal{P}^*}\left(T^1 > t^*\right) - P_{\cal{P}^*}\left(T^0 > t^*\right)\\
	\label{eq-estimand1}
	& E_{\cal{P}^*}\left(Y^1(t^*) | T^1 > t^*\right) - E_{\cal{P}^*}\left(Y^0(t^*) | T^0 > t^*\right)
\end{align}

Survival difference (\ref{eq-estimand2}) is  a typical primary outcome in clinical trials and observational studies alike. A mean difference  among the living (\ref{eq-estimand1}) can be caused by treatment in two ways. Besides having a potential direct impact on $Y$, the treatment affects who dies. Hence, (\ref{eq-estimand1}) compares potentially differently evolved subsets of survivors following either treatment. This is part of the very impact and consequence of treatment choice we seek to reveal. Even though not a direct contrast of individual counterfactuals, it may show that those who survive 3 years on treatment A tend to have a good quality of life, while those surviving 3 years on treatment B tend to have a rather low quality of life. In combination with the survival probabilities, this provides actionable information.

While (\ref{eq-estimand2}) is readily interpreted as an average over individual effects, the different subsets render (\ref{eq-estimand1}) primarily interpretable as a population level causal effect \cite{maldonado}. By defining the multivariate, two-dimensional outcome, similarly to \cite{flanders2015}, (\ref{eq-estimand1}) fits perfectly in Hern\`an and Robin's definition of population causal effects as \textit{a contrast of functionals (including the median, variance, hazard or cdf) of counterfactual outcomes} \cite[ p. ~7]{hernan2020}.

In an ideal world we  would inform patients on individual counterfactual survival time distribution $T_i^a$  and the corresponding $Y$-trajectory $\mathbf{\overline{Y}}_i^a(T_i^{a-})$, allowing them to balance potential survival-benefit with the impact on $Y$ of the treatment choice. In practice, regardless of how intercurrent events are addressed, in clinical trials we estimate average (conditional) effects that cover a range of individual- or subgroup-effects not explicitly estimated. Without more detailed information, these average effects  are what we  know and can act on. Hence, a positive population-level effect may be concluded from an RCT, even when in an RCT the 20\% with an observed positive outcome on standard of care could theoretically be a subset of the 60\% with a negative outcome on the experimental arm.

With  similar impact on individual $Y$ over time, when one treatment keeps more frail patients alive, average $Y(t)$ may be worse among the larger group of survivors. Such scenario, however, would show up in the survival-dimension. Baseline covariate-specific results will also reflect this, if baseline-characteristics like disease-severity or starting point $Y(t=0)$ capture frailty well. It is important here to recognize that some patients may well find longer survival with bad $Y$ undesirable.

\section{Estimation} \label{sec-estimation}

Focusing first on a fixed time $t^*$ coinciding with one of the scheduled $t_j$, we start from the sample as described in Section~\ref{sec-notprel}. To ignore complications with intermittent missingness \cite{thomassen2025}, we assume no missing observations on $\overline{Y}_{i}(t^*)$ except after censoring. Hence, for $Y_i(t)$ or any other time-varying covariate to be measured, we need both $T_i > t$ and $C_i > t$. To estimate the two-dimensional while-alive estimand in target population ${\cal{P}^*}$, we require the following causal identification assumptions \cite{goetghebeur2020}:

\begin{itemize}
	\item[A1.] Consistency: $(T^a, \overline{\mathbf{Y}}^a(T^{a-}))=(T, \overline{\mathbf{Y}}(T^-))$  when $A=a$ 
	\item[A2.] No Interference $(T_i^{a_i}, \overline{\mathbf{Y}}_i^{a_i}(T^{a_i-})) = (T_i^{\mathbf{a}}, \overline{\mathbf{Y}}_i^{\mathbf{a}}(T^{\mathbf{a}-}))$ with $\mathbf{a}$ the vector of counterfactual treatments for all patients.
	\item[A3.] Positivity: $\forall \mathbf{z} \text{ with } f_{\cal{P}*}(\mathbf{z})>0: P(A=a|\mathbf{Z}=\mathbf{z})>0$ for $a=0$ and $a=1$ or as it will be used equivalently $\forall \mathbf{z} \text{ with } f_{\cal{P}*}(\mathbf{z})>0: f_{{\cal{P}}_0}(\mathbf{z})>0  \text{ and }  f_{{\cal{P}}_1}(\mathbf{z})>0$
	\item[A4.] Conditional Exchangeability: $(T^a, \overline{\mathbf{Y}}^a(T^{a-})) \ind A |\mathbf{Z}$
\end{itemize}
In the presence of censoring, additional assumptions are required for identification. Either:
\begin{itemize}		
	\item[A5a.] Conditional Independent Censoring for
	\begin{itemize}
		\item[1.] Survival: $T\ind C|\mathbf{Z}$
		\item[2.] Longitudinal Outcome: $E(Y(t^*)|A, \mathbf{Z}, T>t^*, C > t^*) = E(Y(t^*)|A, \mathbf{Z}, T>t^*)$
	\end{itemize}
\end{itemize}
or with dependent censoring:
\begin{itemize}		
	\item[A5b.] Conditional on measured time-varying covariate history $\overline{\mathbf{V}}(t^-)$ and the outcome history $\overline{\mathbf {Y}}(t^-)$, both up to but excluding $t$ (as $Y_i(t)$ and $V_i(t)$ are only measured if $X_i > t$):
	\begin{itemize}
		\item[1.] Survival: $\lambda_C(t|\overline{\mathbf{V}}(t^-), \overline{\mathbf{Y}}(t^-),A, \mathbf{Z}, T, T>t)=\lambda_C(t|\overline{\mathbf{V}}(t^-), \overline{\mathbf{Y}}(t^-),A, \mathbf{Z}, T>t)$ \cite{robins2000noncomp}
		\item[2.] Longitudinal Outcome \cite{kurland2005, rouanet2019}: $$P(C>t|Y(t),\overline{\mathbf{V}}(t^-),A, \mathbf{Z}, T, T>t) = P(C>t|\overline{\mathbf{Y}}(t^-),\overline{\mathbf{V}}(t^-), A, \mathbf{Z}, T>t)$$ 
	\end{itemize}
\end{itemize}

Under these assumptions, the survival component can be estimated using IPTCW \cite{hernan2000msm} or regression-standardization \cite{makuch1982}.

The treatment-specific terms in (\ref{eq-estimand1}) can be estimated by averaging the $\mathbf{Z}$-specific expected $Y$-while-alive at time $t^*$ over the $\mathbf{Z}$-distribution of those still alive. This $\mathbf{Z}$-distribution weights the baseline densities by their survival probabilities at $t^*$.
%\begin{equation}
\begin{eqnarray}  \phantomsection\label{eq-sep}
	&&	E_{\cal{P}^*}\left(Y^a(t^*) | T^a > t^*\right)  \nonumber\\
	&&	 =  \int_{\cal{P}^*} E_{\cal{P}^*}(Y^a(t^*)|\mathbf{Z}=\mathbf{z}, T^a > t^* ) \cdot f_{\cal{P}^*}(\mathbf{Z}=\mathbf{z}|T^a>t^*)\cdot \dd\mathbf{z}   \label{eq-sepa}\\
	%  &\text{which can be written without counterfactual variables as: } \\
	%=   \hat{E}_{\cal{P}^*}\left(Y^a(t^*) | T^a > t^*\right)
	&&	 =  \frac{\int_{\cal{P}^*} E(Y(t^*)| A = a, \mathbf{Z}=\mathbf{z},  T > t^* ) \cdot P(T>t|A = a, \mathbf{Z}=\mathbf{z}) \cdot f_{\cal{P}^*}(z)\cdot \dd\mathbf{z}}{\int_{\cal{P}^*} P(T>t|A = a, \mathbf{Z}=\mathbf{z}) \cdot f_{\cal{P}^*}(z)\cdot \dd\mathbf{z}}  \label{eq-sepb}
\end{eqnarray}

where the first step applies the law of iterated expectations and the step from the counterfactuals in (3) to the observables in (4) rests on assumptions A1-A4.

Appendix Figure \ref{fig-scheme1b} illustrates how the while-alive estimand is impacted by $\mathbf{Z}$-conditional $Y$-while-alive data and $\mathbf{Z}$-specific survival under both treatments. With differential survival between observed treatments, their exchangeable populations at baseline may generate surviving subpopulations at later times with different baseline-$\mathbf{Z}$ distributions. It is over those different subpopulations that we average $Y$-while-alive.

\subsection{Estimation in the absence of censoring}\label{sec-nocens}

Both methods below rely on exchangeability, positivity, consistency and non-interference, assumptions A1-A4 in Section \ref{sec-estimation}.

\subsubsection{Regression-standardization}\label{sec-regstandnocens}

Estimates of
$E(Y(t^*)| \mathbf{Z}=\mathbf{z}, A=a, T > t^*)$  in (\ref{eq-sepb}), are readily derived from the observed data using e.g. a (flexible) parametric mean model. The second factor, $P(T>t|\mathbf{Z}=\mathbf{z}, A=a)$, follows upon fitting Cox regression or other time-to-event methods. The integral in (\ref{eq-sepb}) can then either involve the known $\mathbf{Z}$-density, or be empirically estimated from a (weighted) sample $S^*$ representing the target population as shown in (\ref{eq-regstnocens}).

\begin{equation}\phantomsection\label{eq-regstnocens}{
		\begin{aligned}
			&\hat{E}_{\cal{P}^*}\left(Y^1(t^*) | T^1 > t^*\right) - \hat{E}_{\cal{P}^*}\left(Y^0(t^*) | T^0 > t^*\right) \\
			&= \frac{\sum_{i \in S^*} \hat{E}(Y(t^*)|A = 1, \mathbf{Z}=\mathbf{z}_i, T > t^*)\cdot \hat{P}(T > t^* | A = 1, \mathbf{Z}=\mathbf{z}_i)}{\sum_{i \in S^*} \hat{P}(T > t^* | A = 1, \mathbf{Z}=\mathbf{z}_i)}\\
			&\hphantom{a+b} - \frac{\sum_{i \in S^*} \hat{E}(Y(t^*)|A = 0, \mathbf{Z}=\mathbf{z}_i, T > t^*) \cdot \hat{P}(T > t^* | A = 0, \mathbf{Z}=\mathbf{z}_i)}{\sum_{i \in S^*} \hat{P}(T > t^* | A = 0, \mathbf{Z}=\mathbf{z}_i)}
		\end{aligned}
}\end{equation}

\subsubsection{IPT Weighting}\label{sec-weightnocens}

In the absence of censoring, those still observed equal those still alive. Hence we can estimate the expected $Y$-while-alive from those observed as
\begin{equation}\phantomsection\label{eq-weightnocens}{
		\begin{aligned}
			E\left(\frac{\sum_{i \in S_a} I(T_i>t^*) \cdot Y_i(t^*)}{\sum_{i \in S_a} I(T_i>t^*)}\right) %\\  
			= E_{\mathcal{P}_a}(Y(t^*) | A = a, T > t^*)
		\end{aligned}
}\end{equation}

Where $I(x)$ is 1 if $x$ is true and 0 if false.

To transport this result towards target populations at baseline, one can use IPT-Weights for each observation in each treatment group: $w^a_{IPTi} = \frac{f_{\cal{P}^*}(z_i)}{f_{S_a}(z_i)}$. The IPT-weighted average outcome among those observed is then:

\begin{equation}\phantomsection\label{eq-weightnocens2}{
		\hat{E}_{\cal{P}^*}\left(Y^a(t^*) | T^a > t^*\right) = \frac{\sum_{i \in S_a} \hat{w}^a_{IPTi}  \cdot I(T_i>t^*) \cdot Y_i(t^*)}{\sum_{i \in S_a}  \hat{w}^a_{IPTi}\cdot I(T_i>t^*)}
}\end{equation}

These weights can be estimated, using for example logistic regression, from a dataset where $S_a$ and $S^*$ are combined (if $S^*$ is defined as e.g. a subset of $S_a$, then these observations are in the combined dataset twice. Once assigned to $S_a$, once to $S^*$). The required weights are then the odds of belonging to $S^*$ ($\hat{w}^a_{IPTi}=\frac{P(S = s^*|\mathbf{Z}=\mathbf{z}_i)}{1-P(S = s^*|\mathbf{Z}=\mathbf{z}_i)}$).

\subsection{With Censoring Present}\label{sec-including-censoring}

When participants' survival time is censored during the study, those observed may no longer be representative of all those still alive. Identification remains possible when either A5a or A5b holds.
%\begin{itemize}

%		\item[A5.] Conditional Independent Censoring for
%		\begin{itemize}
	%			\item[a.] Survival: $T\ind C|\mathbf{Z}$
	%			\item[b.] Longitudinal Outcome: $E(Y(t^*)|A, \mathbf{Z}, T>t^*, C > t^*) = E(Y(t^*)|A, \mathbf{Z}, %T>t^*)$
	%		\end{itemize}
%	\end{itemize}

\subsubsection{Regression Standardization}\label{sec-regstandcens}

Under A5a, censoring is independent from $Y(t)$ and $T$ conditional on $\mathbf{Z}$. Hence $P(T > t^* | A=a, \mathbf{Z}=\mathbf{z})$ can be estimated under standard non-informative censoring conditional on $\mathbf{Z}$ and

\begin{equation}\phantomsection\label{eq-regstand}{
	\begin{aligned}
		E(Y(t^*)|A=a,\mathbf{Z}=\mathbf{z}, T > t^*) & = E(Y(t^*)|A = a, \mathbf{Z}=\mathbf{z}, X > t^*)
	\end{aligned}
}\end{equation}

The remaining calculations follow the approach in \ref{sec-regstandnocens}. The baseline covariate set may need to be broadened beyond the treatment confounders set for this assumption to hold. If we must rely on A5b, then we need to incorporate IPCW-weights properly accounting for the recorded covariate history $\overline{\mathbf{V}}(t^{*-})$ and outcome history $\overline{\mathbf{Y}}_i(t^{*-})$. 

\subsubsection{IPTC Weighting}\label{sec-weightcens}

To translate the sample of those observed ($X_i > t^*$) to a sample of those alive ($T_i>t^*$), inverse probability of censoring weighting can be applied with IPC Weights $w^a_{IPCi}(t^*) = P(C > t^* |\overline{\mathbf{V}}(t^{*-})=\overline{\mathbf{v}}_i(t^{*-}), \overline{\mathbf{Y}}(t^{*-})=\overline{\mathbf{y}}_i(t^{*-}),\mathbf{Z}=\mathbf{z}_i,A = a, T_i>t^*)^{-1}$ \cite{kurland2005, rouanet2019}, where $\overline{\mathbf{V}}(t^{*-})$ can be ignored if A5a holds instead of A5b. Therefore, 
multiplying treatment balancing weights in (\ref{eq-weightnocens2}) 
with these IPC weights, arrives at an  Inverse Probability of Treatment and Censoring Weighted (IPTCW) estimator:

\begin{equation}\phantomsection\label{eq-weightcenstot}{
	\hat{E}_{\cal{P}^*}\left(Y^a(t^*) | T^a > t^*\right) = \frac{\sum_{i \in S_a} \hat{w}^a_{IPCi}(t^*)\cdot \hat{w}^a_{IPTi}  \cdot I(T_i > t^*, C_i > t^*) \cdot y_i}{\sum_{i \in S_a} \hat{w}^a_{IPCi}(t^*)\cdot \hat{w}^a_{IPTi}\cdot I(T_i > t^*, C_i > t^*)}
}\end{equation}

Standard time-to-event methods like Cox-regression can be applied within each treatment-cohort, to estimate these IPC-weights by defining 'being censored' as the event of interest among those observed.

\subsection{Longitudinal Analysis}\label{sec-long}

To move beyond a specific time point \(t^*\), we perform a longitudinal, joint analysis over all time points. Kurland and Heagerty \cite{kurland2005} developed an IPC-Weighted GEE-approach with an independence working correlation (IPCW-IEE) to avoid the implicit imputation of outcomes after death. This can be employed as input for the regression-standardization approach in (\ref{eq-regstnocens}). Otherwise, adding an IPTW-step to their approach for 'Regression Conditional on being Alive', makes for a longitudinal version of the presented time-specific IPTCW-approach (\ref{eq-weightcenstot}). For this and previous approaches, the bootstrap can be used to estimate uncertainty.

\section{Simulation} \label{sec-sim}

We illustrate finite sample performance of the methods presented above using a simple simulation with binary actionable treatment \(A\), one continuous baseline confounder \(Z\), longitudinal continuous outcome \(Y(t)\) generated with a person-specific random intercept and evaluated at at baseline $t_0=0$ and at 10 planned follow up times with $t_j=j \in \{1, 2,\ldots,10\}$. Survival time \(T\) and censoring time \(C\) are evaluated in continuous time. 

The presented scenario highlights the potential impact of ignoring confounding and dependent censoring. Those treated (\(A=1\)), tend to have higher \(Z\) and irrespective of the treatment received, higher \(Z\) is associated with higher \(Y(0)\). Over time, \(Y\) tends to increase within individuals and more so if \(A=1\) and with higher \(Z\) conditional on $A$. Over time t, the hazards for death and censoring depend on baseline A, Z and $Y(t_m)$ where $t_m$ is the most recent $t_j$ preceding $t$. For ease of expression, we will write $Y(\lfloor t \rfloor)$. For \(A=0\), the hazards for death and censoring decrease with increasing $Y(\lfloor t \rfloor)$. For \(A=1\), this is reversed. Full detail on the data generating mechanism  can be found in Appendix B. R-code is available at https://github.com/DRUgent/WhileAlive.

As naive IEE methods in this setting we estimate 1) the difference in observed average $Y$ (\textit{As Observed}), 2) the difference in IPC (censoring model including $Y(\lfloor t \rfloor)$ as time varying covariate) weighted average $Y$ (\textit{IPCW only}) and 3) difference in IPT weighted average $Y$ (\textit{IPTW only}). 

\textit{IEE-IPTCW} and IEE-Regression Standardization (\textit{IEE-RegStand}) are presented first naively, assuming censoring depends only on baseline-covariates, and then allowing $Y(\lfloor t \rfloor)$ to influence the time-specific hazard for censoring. %For the regression standardization, this implies that in the second case, both the $Y-$ and survival-model are IPC weighted.

\begin{figure}
%\centering{
	\includegraphics[width=0.92\textwidth, height=0.92\textheight]{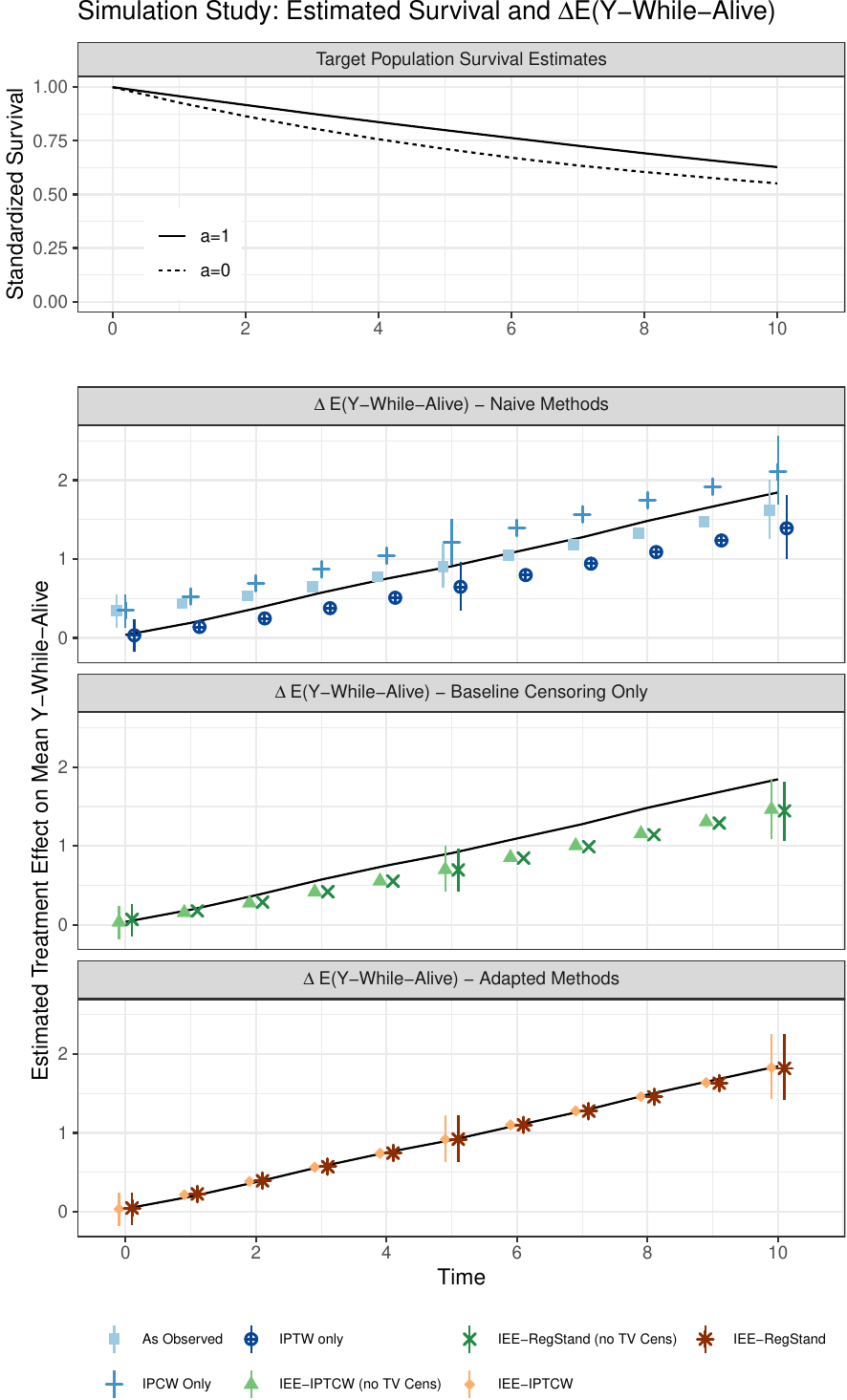}
	%}
\caption{\label{fig-simres-1} The top panel provides target population survival estimates. The other panels show estimated and true differences in mean Y-while-alive over time by method (solid line represent the true values, calculated from a large N dataset). CIs are presented for selected time points only.}

\end{figure}

Results are presented in Figure \ref{fig-simres-1} and Appendix Table \ref{tbl-bias}. In our scenario,
IPCW-only, with uncorrected baseline confounding, a positive bias emerges for the while-alive-treatment effect because with lower \(Z\) for $A=0$ also comes lower \(Y\). This positive bias remains high over time. IPTW-only without accounting for censoring, artificially improves the results for \(A=0\) (where mainly low \(Y\)-patients are censored) and the opposite for \(A = 1\). Hence a negative bias for the treatment effect that grows over time with increasing censoring. The as-observed analysis lies in between these two extremes, starting close to IPC, but ending up closer to IPTW. It is also biased.

The methods that account for both confounding and censoring, perform much better when the impact of $Y(t)$ on the censoring is properly accounted for. The coverage of the percentile bootstrap-based confidence intervals is close to the nominal 0.95 (Appendix Table \ref{tbl-bias}). In this simulation, IPTCW and Regression Standardization have similar empirical standard deviations.% In the next section, we apply these methods to a case study presented in the context of SISAQOL-IMI  (Setting International Standards in Analysing Patient-Reported Outcomes and Quality of Life Endpoints in Cancer Clinical Trials ? IMI) \cite{pe2023}.

\section{Case study: Health Related QoL in Single Arm Trial vs External control} \label{sec-casestudy}

\subsection{Setting}

We apply the methodology for the two-dimensional while-alive estimand, to estimate the impact of an experimental treatment on Health Related QoL in patients with late stage cancer in a Single Arm Trial (SAT) with external control. Such SATs play an increasingly important role in cancer clinical research.  With RCTs sometimes deemed impractical or unethical, a growing number of FDA and EMA approvals rest on pivotal evidence of SATs \cite{ReflectionPaperEstablishingb, hilal2020}.

Without concurrent randomized controls, causal inference helps drawing comparative effectiveness conclusions under well understood assumptions, which may be questioned and be subject to sensitivity analyses in a second stage. The question we study below was put forward and discussed in a large consortium of stakeholders collaborating in the context of SISAQOL-IMI  (Setting International Standards in Analysing Patient-Reported Outcomes and Quality of Life Endpoints in Cancer Clinical Trials) \cite{pe2023, amdal2025}.

\subsection{Data sources}

We analyse EORTC QLQ-C30 global QoL scale measures \cite{aaronson1993}, a secondary outcome in an SAT of patients with non-small cell lung cancer \cite{blackhall2014, blackhall2017}. 

The same experimental treatment was also studied in an RCT with chemotherapy as control \cite{shaw2013} using EORTC QLQ-C30 under an identical three-weekly measurement schedule. This control arm is a suitable comparator for the SAT, notwithstanding the RCT was restricted to patients with fewer comorbidities and fewer prior lines of treatment. Additionally, the experimental RCT-arm can serve as a negative control \cite{lipsitch2010}. With identical treatment under similar settings, we expect no structural differences in outcomes with the SAT after adjustment for confounding in overlapping (sub)populations. Contradicting results would indicate that the adjustment for confounding was insufficient or that other issues (e.g. informative drop out, differences in surrounding care) were not properly addressed.

For our purpose here, we start from two datasets, anonymized by the sponsor, with single imputation for intermittent missingness. QoL was imputed until time of death or censoring, using time-varying information on treatment discontinuation, progression and time to death as described in \cite{thomassen2024}. This allows our analysis to focus on the two-dimensional while-alive estimand, confounding and censoring.

\subsection{Analysis}

We assume age, sex, the number of prior treatments, ECOG performance status and 
QoL at baseline  form a sufficient set of baseline confounders for treatment exchangeability and also to render censoring non-informative (methods labelled BC in Figure \ref{fig-treateff}). In a second instance we will allow censoring to further depend on the observed (or imputed) QoL over time and an indicator of disease progression (methods labelled TVC in Figure \ref{fig-treateff}).  The target population for treatment is best represented by the SAT population here, which is our initial choice. We will further reduce the scope, however, due to positivity constraints. For sufficient overlap in the considered baseline confounders, the target population is restricted to the SAT-sub-population with ECOG \(\leq\) 2 and fewer than 3 prior drug therapies  (Appendix Tables \ref{tbl-demcomp} and \ref{tbl-demred})).

Treatment group differences in mean QoL-while-alive  are estimated at three-week intervals using the methods presented in sections \ref{sec-including-censoring} and \ref{sec-long},  with bootstrap based  confidence intervals.  The standardized survival curves were derived from Cox regression models \cite{andersen2021}, fitting separate Cox models  per treatment before  averaging the resulting predicted survival curves over the common covariate distribution of the target population. This was performed for the primary analysis comparing the experimental treatment in the SAT with the control arm in the RCT and for the sensitivity (negative control) analysis, comparing the experimental treatment groups in SAT and RCT.

\subsection{Results}

\begin{figure}
%\centering{
	\includegraphics[width=0.92\textwidth, height=0.92\textheight]{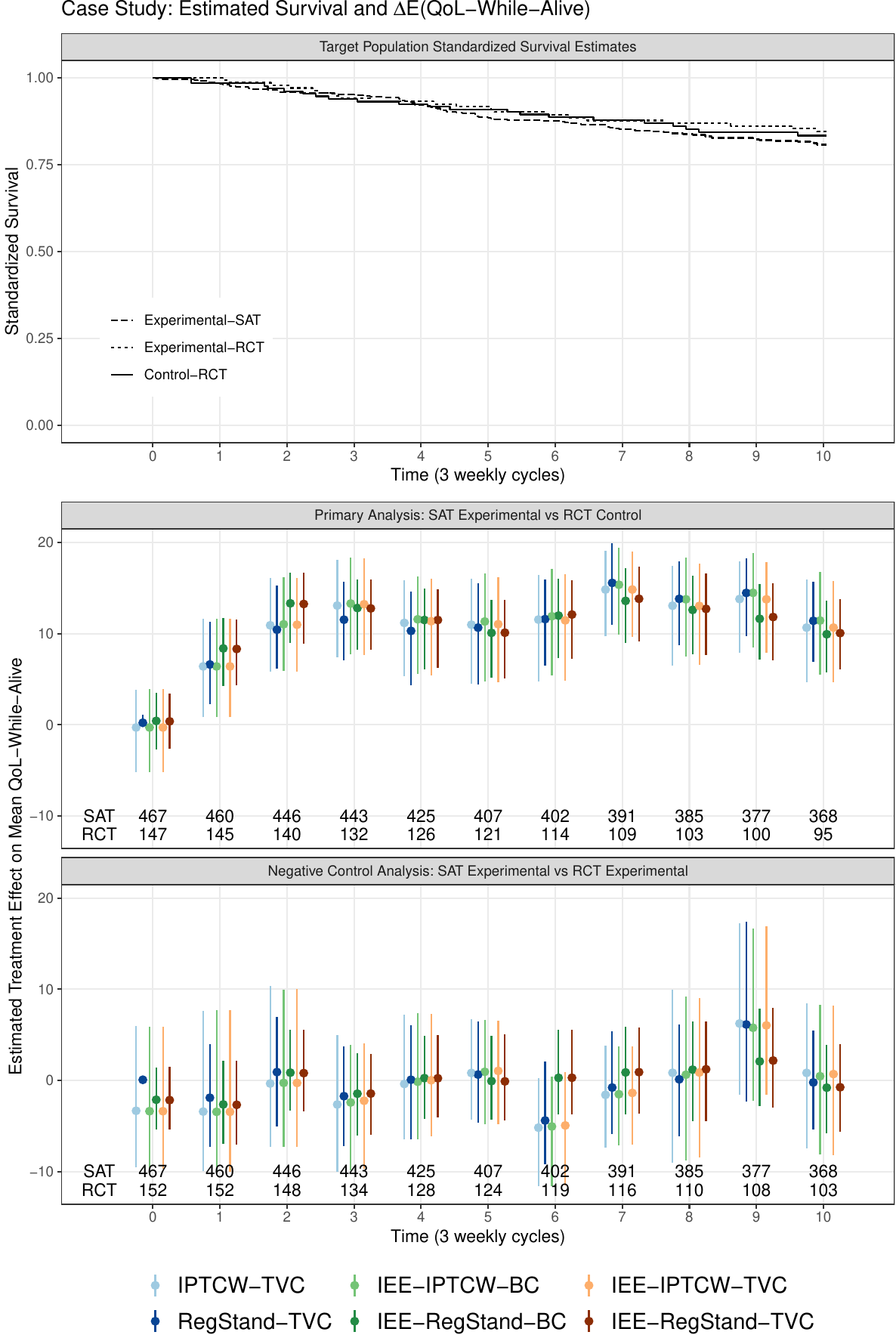}
	%}
\caption{\label{fig-treateff} Estimated differences in Qol-while-alive between SAT and the respective arms of the RCT (Top panel: Standardized survival curves. Middle panel: main analysis for efficacy. Dots are the point estimates, vertical lines represent bootstrapped 95\% confidence intervals. Bottom panel: negative control analysis comparing SAT with the experimental RCT-arm). Outcomes labelled with/without 'IEE' use longitudinal/single time point-analyses. 'TVC' indicate the censoring model incorporates time-varying covariates, 'BC' only baseline covariates }

\end{figure}

The upper panel of Figure \ref{fig-treateff} shows nearly identical survival curves in the SAT and the RCT control arm, while the middle panel  reveals  better mean Qol-while-alive in the SAT up until week 30 (10 cycles). These QoL-results hardly vary over the specific estimation methods shown, indicating limited effect of the specific censoring mechanism considered in this case.  
Importantly, the comparable survival outcomes show the favourable QoL-result is not a mere selection-through-survival effect. 

The narrow confidence interval at $t=0$ for the time point specific regression-standardization in Figure \ref{fig-treateff} is an artefact following inclusion of  baseline QoL in the regression model. The baseline comparisons do reveal how the other methods manage to balance QoL at baseline between treatment arms. At later time points,  IPTCW-methods have wider confidence intervals than their regression-standardization counterparts, as expected.

In the sensitivity analysis ('negative control analysis') where the experimental SAT group was compared  with the experimental RCT-arm, estimated differences in QoL while-alive are mostly close to zero with CIs overlapping zero. This is consistent with the 'no unmeasured confounders' assumption.

\section{Discussion} \label{sec-disc}

In this paper, we advocated the use of the two-dimensional outcome $Y_{2D}$ consisting of survival and longitudinal repeated measures, with a corresponding two-dimensional while-alive estimand. The estimand provides a meaningful target for the total effect of treatment. It is relevant in randomized and observational settings, and the proposed estimation approaches are similarly applicable. Within well conducted RCTs, exchangeability and consistency are generally ensured by design, but positivity may be violated if the target population differs from the RCT-population. Censoring, is an important consideration in both settings.

Of course, not all causal effects are of equal interest. Sj\"olander \cite{sjolander2011} argues that '\textit{any causal effect deserves attention if it can improve decision making [$\ldots$] or it generates insights about the problem at hand (e.g. about etiology or biological mechanisms).}' The proposed estimand is less focused on etiology, but critically informs decision making. Targeting both the contrast in survival (\ref{eq-estimand2}) and in $Y$-while-alive (\ref{eq-estimand1}), allows the actually realized $Y$ to be acknowledged on its own and be balanced against survival. This explicit evaluation  of the survival component differs from alternative, popular approaches to handle death. Combining longitudinal outcomes with death into a composite outcome, focusing on hypothetical scenarios as if death did not occur, or restriction to those who are guaranteed to survive on both treatments, all result in  estimands that either do not fully distinguish quality of life from not being alive, or target a world that does not inform all patients (nor an observable subset). The causal two-dimensional while-alive estimand addresses distinct and clinically relevant questions.

Our case study found similar survival under both treatments and higher average QoL-while-alive on the experimental treatment. Under the assumptions made, these findings support a favourable assessment of the intervention. As with any marginal effect, including those based on randomized trials, differences in average QoL-while-alive may reflect complex underlying mechanisms and will average over a range of possibly differing effects across sub-cohorts. For QoL-while-alive, however, the contrast depends on treatment impact on both survival and QoL. If the estimated treatment effects  go in seemingly opposite directions for the two dimensions, the two-dimensional estimand still highlights this tangible reality, which requires decision making that is less straightforward. Some patients may prefer the prospect of improved survival, others prioritize quality of life over length of life. For individual patients, stratified $\mathbf{Z}$-specific estimates may provide additional insights, to help make these decisions. 

The proposed estimation approaches build on existing work on partly-conditional regression \cite{kurland2005, dufouil2004, wen2018a}. Methods for handling missingness as presented in \cite{wen2018a, wen2018b} can be incorporated. The simulation study showed well behaved small sample properties of our estimators under the stated assumptions. The estimators share a set of causal identification assumptions to account for confounding and for non-informative censoring,  possibly relying on time-varying characteristics. While they model either treatment propensity or outcome, both methods provided similar estimates. In the case study, smaller confidence intervals for the regression-standardization approaches, reflect their ability to exploit their modelling assumptions if they hold (e.g. linear or smooth trends, lack of interactions,...) and to then extrapolate estimates to profiles that completely dropped out of the study at $t^*$. By building on existing methods, estimation can be implemented combining available software.

As many of our assumptions are untestable on the observed data, sensitivity analyses \cite{dang2023} are an inherent part of any causal analysis. Here, the case study lends itself perfectly to a negative control analysis. Indeed, if all assumptions are reasonably met, we would not expect systematic differences in outcomes between the experimentally treated in the SAT and in the RCT, after adjustment for known baseline confounding and accounting for censoring. In our case, results were remarkably consistent. If, however, this sensitivity analysis would show clear differences, it would point to shortcomings in the assumptions made or the study design.

On top of the causal assumptions, the censoring assumptions are a key component of the analysis. Censoring can depend heavily on time-varying characteristics. This information then should be accounted for to avoid bias, at the cost of increased complexity and more stringent positivity assumptions. In the case study, patients had very limited follow up for QoL after progression or treatment discontinuation. This means that post-progression and post-discontinuation estimates are based on limited available information, complicating a thorough analysis of potentially time-varying covariate-dependent censoring. When feasible, a study design and data collection tailored to the targeted estimand, may mitigate this problem.

The presented bootstrap-based confidence intervals account for uncertainty arising from estimating the weights and the expected outcomes. The choice of external control study is an additional source of heterogeneity-induced uncertainty  that one would wish to account for \cite{collignon2020}. The between-study variability may result from a wide range of sources, both baseline patient mix and post-baseline features (such as surrounding care, measurement tools and schedules, adherence,...). Meta-analysis may provide an estimate for this between-study variability. In our case study,  by selecting studies with similar protocols and measurement tools, some of these sources are accounted for by design. Since our analyses explicitly adjust for case-mix, the between-study variability estimated from an unconditional meta-analysis, may be seen as an upper bound for this additional uncertainty on the estimated treatment effect.

In this paper, we did not discuss joint models, that have been designed to link the survival and longitudinal data model through a shared person-specific random effect. Their fixed effects parameters target estimands in an immortal population \cite{rouanet2019}, but while-alive estimands can be derived by standardizing survival-weighted $Y(t)$ as one averages over the random effects. In a simulation study, \cite{vdp2025} found this to be less efficient than the proposed IEE-methods for this purpose. Their role for prediction of the two-dimensional outcome, is studied elsewhere \cite{vanderpijpen}.

To conclude: the complexities arising in cohorts with considerable mortality, should not stop us from asking fundamentally relevant questions that inform all patients on what they may expect from their treatment choice. While-alive causal estimands are often a key component of this evidence, with their estimation relying on fewer assumptions than most alternatives.

\bibliographystyle{unsrtnat}
\bibliography{references}

@article{kurland2005,
	title = {Directly parameterized regression conditioning on being alive: analysis of longitudinal data truncated by deaths},
	author = {Kurland, Brenda F. and Heagerty, Patrick J.},
	year = {2005},
	month = {04},
	date = {2005-04-01},
	journal = {Biostatistics},
	pages = {241--258},
	volume = {6},
	number = {2},
	doi = {10.1093/biostatistics/kxi006}
}

@article{shardell2015,
	title = {Doubly robust estimation and causal inference in longitudinal studies with dropout and truncation by death},
	author = {Shardell, Michelle and Hicks, Gregory E. and Ferrucci, Luigi},
	year = {2015},
	month = {01},
	date = {2015-01-01},
	journal = {Biostatistics},
	pages = {155--168},
	volume = {16},
	number = {1},
	doi = {10.1093/biostatistics/kxu032}
}

@article{tchetgentchetgen2012,
	title = {Rejoinder: To Weight or Not to Weight?: On the Relation Between Inverse-probability Weighting and Principal Stratification for Truncation by Death},
	author = {Tchetgen Tchetgen, Eric J. and Glymour, M. Maria and Shpitser, Ilya and Weuve, Jennifer},
	year = {2012},
	month = {01},
	date = {2012-01},
	journal = {Epidemiology},
	pages = {132},
	volume = {23},
	number = {1},
	doi = {10.1097/EDE.0b013e31823b5081}
}

@article{kurland2009,
	title = {Longitudinal Data with Follow-up Truncated by Death: Match the Analysis Method to Research Aims},
	author = {Kurland, Brenda F. and Johnson, Laura L. and Egleston, Brian L. and Diehr, Paula H.},
	year = {2009},
	date = {2009},
	journal = {Statistical science : a review journal of the Institute of Mathematical Statistics},
	pages = {211},
	volume = {24},
	number = {2},
	doi = {10.1214/09-STS293}
}

@article{blackhall2017,
	title = {Final results of the large-scale multinational trial PROFILE 1005: efficacy and safety of crizotinib in previously treated patients with advanced/metastatic ALK-positive non-small-cell lung cancer},
	author = {Blackhall, Fiona and Ross Camidge, D. and Shaw, Alice T. and Soria, Jean-Charles and Solomon, Benjamin J. and Mok, Tony and Hirsh, Vera and {Jänne}, Pasi A. and Shi, Yuankai and Yang, Pan-Chyr and De Pas, Tommaso and Hida, Toyoaki and {De Carpeño}, Javier Castro and Lanzalone, Silvana and Polli, Anna and Iyer, Shrividya and Reisman, Arlene and Wilner, Keith D. and Kim, Dong-Wan},
	year = {2017},
	month = {01},
	date = {2017-01-01},
	journal = {ESMO Open},
	pages = {e000219},
	volume = {2},
	number = {3},
	doi = {10.1136/esmoopen-2017-000219}
}

@article{blackhall2014,
	title = {Patient-Reported Outcomes and Quality of Life in PROFILE 1007: A Randomized Trial of Crizotinib Compared with Chemotherapy in Previously Treated Patients with ALK-Positive Advanced Non{\textendash}Small-Cell Lung Cancer},
	author = {Blackhall, Fiona and Kim, Dong-Wan and Besse, Benjamin and Nokihara, Hiroshi and Han, Ji-Youn and Wilner, Keith D. and Reisman, Arlene and Iyer, Shrividya and Hirsh, Vera and Shaw, Alice T.},
	year = {2014},
	month = {11},
	date = {2014-11-01},
	journal = {Journal of Thoracic Oncology},
	pages = {1625--1633},
	volume = {9},
	number = {11},
	doi = {10.1097/JTO.0000000000000318}
}

@article{aaronson1993,
	title = {The European Organization for Research and Treatment of Cancer QLQ-C30: A Quality-of-Life Instrument for Use in International Clinical Trials in Oncology},
	author = {Aaronson, Neil K. and Ahmedzai, Sam and Bergman, Bengt and Bullinger, Monika and Cull, Ann and Duez, Nicole J. and Filiberti, Antonio and Flechtner, Henning and Fleishman, Stewart B. and Haes, Johanna C. J. M. de and Kaasa, Stein and Klee, Marianne and Osoba, David and Razavi, Darius and Rofe, Peter B. and Schraub, Simon and Sneeuw, Kommer and Sullivan, Marianne and Takeda, Fumikazu},
	year = {1993},
	month = {03},
	date = {1993-03-03},
	journal = {JNCI: Journal of the National Cancer Institute},
	pages = {365--376},
	volume = {85},
	number = {5},
	doi = {10.1093/jnci/85.5.365}
}

@article{hilal2020,
	title = {Limitations in Clinical Trials Leading to Anticancer Drug Approvals by the US Food and Drug Administration},
	author = {Hilal, Talal and Gonzalez-Velez, Miguel and Prasad, Vinay},
	year = {2020},
	month = {08},
	date = {2020-08-01},
	journal = {JAMA internal medicine},
	pages = {1108--1115},
	volume = {180},
	number = {8},
	doi = {10.1001/jamainternmed.2020.2250}
}

@article{goetghebeur2020,
	title = {Formulating causal questions and principled statistical answers},
	author = {Goetghebeur, Els and le Cessie, Saskia and De Stavola, Bianca and Moodie, Erica EM and Waernbaum, Ingeborg},
	year = {2020},
	date = {2020},
	journal = {Statistics in Medicine},
	pages = {4922--4948},
	volume = {39},
	number = {30},
	doi = {10.1002/sim.8741}
}

@article{thomassen2024,
	title = {The role of the estimand framework in the analysis of patient-reported outcomes in single-arm trials: a case study in oncology},
	author = {Thomassen, Doranne and Roychoudhury, Satrajit and Amdal, Cecilie Delphin and Reynders, Dries and Musoro, Jammbe Z. and Sauerbrei, Willi and Goetghebeur, Els and le Cessie, Saskia and others},
	year = {2024},
	month = {11},
	date = {2024-11-23},
	journal = {BMC Medical Research Methodology},
	pages = {290},
	volume = {24},
	number = {1},
	doi = {10.1186/s12874-024-02408-x}
}

@article{shaw2013,
	title = {Crizotinib versus Chemotherapy in Advanced ALK-Positive Lung Cancer},
	author = {Shaw, Alice T. and Kim, Dong-Wan and Nakagawa, Kazuhiko and Seto, Takashi and {Crinó}, Lucio and Ahn, Myung-Ju and De Pas, Tommaso and Besse, Benjamin and Solomon, Benjamin J. and Blackhall, Fiona and Wu, Yi-Long and Thomas, Michael and {O'Byrne}, Kenneth J. and Moro-Sibilot, Denis and Camidge, D. Ross and Mok, Tony and Hirsh, Vera and Riely, Gregory J. and Iyer, Shrividya and Tassell, Vanessa and Polli, Anna and Wilner, Keith D. and {Jänne}, Pasi A.},
	year = {2013},
	month = {06},
	date = {2013-06-20},
	journal = {New England Journal of Medicine},
	pages = {2385--2394},
	volume = {368},
	number = {25},
	doi = {10.1056/NEJMoa1214886}
}

@article{rouanet2019,
	title = {Interpretation of mixed models and marginal models with cohort attrition due to death and drop-out},
	author = {Rouanet, {Anaïs} and Helmer, Catherine and Dartigues, {Jean-François} and Jacqmin-Gadda, {Hélène}},
	year = {2019},
	month = {02},
	date = {2019-02-01},
	journal = {Statistical Methods in Medical Research},
	pages = {343--356},
	volume = {28},
	number = {2},
	doi = {10.1177/0962280217723675}
}

@article{wang2017,
	title = {Inference in randomized trials with death and missingness},
	author = {Wang, Chenguang and Scharfstein, Daniel O. and Colantuoni, Elizabeth and Girard, Timothy D. and Yan, Ying},
	year = {2017},
	date = {2017},
	journal = {Biometrics},
	pages = {431--440},
	volume = {73},
	number = {2},
	doi = {10.1111/biom.12594}
}

@article{kahan2024,
	title = {The estimands framework: a primer on the ICH E9(R1) addendum},
	author = {Kahan, Brennan C. and Hindley, Joanna and Edwards, Mark and Cro, Suzie and Morris, Tim P.},
	year = {2024},
	month = {01},
	date = {2024-01-23},
	journal = {BMJ},
	pages = {e076316},
	volume = {384},
	doi = {10.1136/bmj-2023-076316}
}

@misc{ReflectionPaperEstablishingb,
	title = {Reflection Paper on Establishing Efficacy Based on Single-Arm Trials Submitted as Pivotal Evidence in a Marketing Authorisation Application},
	author = {{European Medicines Agency}},
	year = {2024},
	url = {https://www.ema.europa.eu/en/documents/scientific-guideline/reflection-paper-establishing-efficacy-based-single-arm-trials-submitted-pivotal-evidence-marketing-authorisation-application_en.pdf}
}

@article{robins1986,
	title = {A new approach to causal inference in mortality studies with a sustained exposure period{\textemdash}application to control of the healthy worker survivor effect},
	author = {Robins, James},
	year = {1986},
	month = {01},
	date = {1986-01-01},
	journal = {Mathematical Modelling},
	pages = {1393--1512},
	volume = {7},
	number = {9},
	doi = {10.1016/0270-0255(86)90088-6}
}

@article{robins2000,
	title = {Marginal Structural Models and Causal Inference in Epidemiology},
	author = {Robins, James M. and {Hernán}, {Miguel Ángel} and Brumback, Babette},
	year = {2000},
	month = {09},
	date = {2000-09},
	journal = {Epidemiology},
	pages = {550},
	volume = {11},
	number = {5},
	doi = {10.1097/00001648-200009000-00011}
}

@article{pe2023,
	title = {Setting International Standards in Analyzing Patient-Reported Outcomes and Quality of Life Endpoints in Cancer Clinical Trials-Innovative Medicines Initiative (SISAQOL-IMI): stakeholder views, objectives, and procedures},
	author = {Pe, Madeline and Alanya, Ahu and Falk, Ragnhild Sorum and Amdal, Cecilie Delphin and Bjordal, Kristin and others},
	year = {2023},
	month = {06},
	date = {2023-06-01},
	journal = {The Lancet Oncology},
	pages = {e270--e283},
	volume = {24},
	number = {6},
	doi = {10.1016/S1470-2045(23)00157-2}
}

@article{vanderweele2011,
	title = {Principal Stratification {\textemdash} Uses and Limitations},
	author = {VanderWeele, Tyler J},
	year = {2011},
	month = {07},
	date = {2011-07-11},
	journal = {The International Journal of Biostatistics},
	pages = {28},
	volume = {7},
	number = {1},
	doi = {10.2202/1557-4679.1329}
}

@article{collignon2020,
	title = {Clustered allocation as a way of understanding historical controls: Components of variation and regulatory considerations},
	author = {Collignon, Olivier and Schritz, Anna and Senn, Stephen J and Spezia, Riccardo},
	year = {2020},
	month = {07},
	date = {2020-07-01},
	journal = {Statistical Methods in Medical Research},
	pages = {1960--1971},
	volume = {29},
	number = {7},
	doi = {10.1177/0962280219880213}
}

@article{egleston2007,
	title = {Causal inference for non-mortality outcomes in the presence of death},
	author = {Egleston, Brian L. and Scharfstein, Daniel O. and Freeman, Ellen E. and West, Sheila K.},
	year = {2007},
	month = {07},
	date = {2007-07-01},
	journal = {Biostatistics},
	pages = {526--545},
	volume = {8},
	number = {3},
	doi = {10.1093/biostatistics/kxl027}
}

@book{hernan2020,
	title = {Causal Inference: What If},
	author = {{Hernán}, Miguel A and Robins, James M},
	year = {2020},
	date = {2020},
	publisher = {CRC Press},
	langid = {en}
}

@article{colantuoni2018,
	title = {Statistical methods to compare functional outcomes in randomized controlled trials with high mortality},
	author = {Colantuoni, Elizabeth and Scharfstein, Daniel O. and Wang, Chenguang and Hashem, Mohamed D. and Leroux, Andrew and Needham, Dale M. and Girard, Timothy D.},
	year = {2018},
	month = {01},
	date = {2018-01-03},
	journal = {BMJ},
	pages = {j5748},
	volume = {360},
	doi={10.1136/bmj.j5748}
}

@misc{europeanmedicinesagency2020,
	title = {{ICH E9 (R1)} addendum on estimands and sensitivity analysis in clinical trials to the guideline on statistical principles for clinical trials},
	author = {{European Medicines Agency}},
	year = {2020},
	date = {2020},
	url = {https://www.ema.europa.eu/en/documents/scientific-guideline/ich-e9-r1-addendum-estimands-sensitivity-analysis-clinical-trials-guideline-statistical-principles_en.pdf.}
}

@article{luo2023,
	title = {Causal inference with outcomes truncated by death in multiarm studies},
	author = {Luo, Shanshan and Li, Wei and He, Yangbo},
	year = {2023},
	date = {2023},
	journal = {Biometrics},
	pages = {502--513},
	volume = {79},
	number = {1},
	doi = {10.1111/biom.13554}
}

@article{dang2023,
	title = {A causal roadmap for generating high-quality real-world evidence},
	author = {Dang, Lauren E. and Gruber, Susan and Lee, Hana and Dahabreh, Issa J. and Stuart, Elizabeth A. and Williamson, Brian D. and Wyss, Richard and {Díaz}, {Iván} and Ghosh, Debashis and {K{\i}c{\i}man}, Emre and Alemayehu, Demissie and Hoffman, Katherine L. and Vossen, Carla Y. and Huml, Raymond A. and Ravn, Henrik and Kvist, Kajsa and Pratley, Richard and Shih, Mei-Chiung and Pennello, Gene and Martin, David and Waddy, Salina P. and Barr, Charles E. and Akacha, Mouna and Buse, John B. and Laan, Mark van der and Petersen, Maya},
	year = {2023},
	month = {01},
	date = {2023-01},
	journal = {Journal of Clinical and Translational Science},
	pages = {e212},
	volume = {7},
	number = {1},
	doi = {10.1017/cts.2023.635}
}

@article{lipsitch2010,
	title = {Negative Controls: A Tool for Detecting Confounding and Bias in Observational Studies},
	author = {Lipsitch, Marc and Tchetgen, Eric Tchetgen and Cohen, Ted},
	year = {2010},
	month = {05},
	date = {2010-05},
	journal = {Epidemiology (Cambridge, Mass.)},
	pages = {383--388},
	volume = {21},
	number = {3},
	doi = {10.1097/EDE.0b013e3181d61eeb}
}

@article{andersen2021,
	title = {Analysis of time-to-event for observational studies: Guidance to the use of intensity models},
	author = {Andersen, Per Kragh and Perme, Maja Pohar and Houwelingen, Hans C. van and Cook, Richard J. and Joly, Pierre and Martinussen, Torben and Taylor, Jeremy M. G. and Abrahamowicz, Michal and Therneau, Terry M.},
	year = {2021},
	date = {2021},
	journal = {Statistics in Medicine},
	pages = {185--211},
	volume = {40},
	number = {1},
	doi = {10.1002/sim.8757}
}

@article{flanders2015,
	title = {A General, Multivariate Definition of Causal Effects in Epidemiology},
	author = {Flanders, W. Dana and Klein, Mitchel},
	year = {2015},
	month = {07},
	date = {2015-07},
	journal = {Epidemiology},
	pages = {481},
	volume = {26},
	number = {4},
	doi = {10.1097/EDE.0000000000000286}
}

@article{maldonado,
	title = {Estimating causal effects},
	author = {Maldonado, George and Greenland, Sander},
	langid = {en},
    journal = {International Journal of Epidemiology},
    year = {2002},
    month = {04},
    date = {2002-04-01},
    volume = {31},
    doi = {10.1093/intjepid/31.2.422}
}

@article{sjolander2011,
	title = {Reaction to Pearl's Critique of Principal Stratification},
	author = {Sjolander, Arvid},
	year = {2011},
	month = {04},
	date = {2011-04-13},
	journal = {The International Journal of Biostatistics},
	volume = {7},
	number = {1},
	doi = {10.2202/1557-4679.1324}
}

@article{makuch1982,
	title = {Adjusted survival curve estimation using covariates},
	author = {Makuch, Robert W.},
	year = {1982},
	month = {01},
	date = {1982-01-01},
	journal = {Journal of Chronic Diseases},
	pages = {437--443},
	volume = {35},
	number = {6},
	doi = {10.1016/0021-9681(82)90058-3}
}

@article{hernan2000msm,
	title = {Marginal Structural Models to Estimate the Causal Effect of Zidovudine on the Survival of HIV-Positive Men},
	author = {{Hernán}, {Miguel Ángel} and Brumback, Babette and Robins, James M.},
	year = {2000},
	month = {09},
	date = {2000-09},
	journal = {Epidemiology},
	pages = {561},
	volume = {11},
	number = {5}
}

@article{robins2000noncomp,
	title = {Correcting for Noncompliance and Dependent Censoring in an AIDS Clinical Trial with Inverse Probability of Censoring Weighted (IPCW) Log{-}Rank Tests},
	author = {Robins, James M. and Finkelstein, Dianne M.},
	year = {2000},
	month = {09},
	date = {2000-09},
	journal = {Biometrics},
	pages = {779--788},
	volume = {56},
	number = {3},
	doi = {10.1111/j.0006-341X.2000.00779.x}
}

@mastersthesis{vdp2025,
    author = {Vanderpijpen, Eline},
    title = {Joint models for expected quality of life: comparison with alternative estimation methods and improved predictions of extreme random effects},
    school = {Ghent University},
    year = {2025}
}

@article{thomassen2025,
	title = {Handling missing values in patient-reported outcome data in the presence of intercurrent events},
	author = {Thomassen, Doranne and Roychoudhury, Satrajit and Amdal, Cecilie Delphin and Reynders, Dries and Musoro, Jammbe Z. and Sauerbrei, Willi and Goetghebeur, Els and le Cessie, Saskia, {on behalf of SISAQOL-IMI Work Package 3} },
	year = {2025},
	month = {03},
	date = {2025-03-01},
	journal = {BMC Medical Research Methodology},
	pages = {56},
	volume = {25},
	number = {1},
	doi = {10.1186/s12874-025-02510-8}
}

@article{amdal2025,
	title = {SISAQOL-IMI consensus-based guidelines to design, analyse, interpret, and present patient-reported outcomes in cancer clinical trials},
	author = {Amdal, Cecilie D. and Falk, Ragnhild S. and Alanya, Ahu and Schlichting, Michael and Roychoudhury, Satrajit and Bhatnagar, Vishal and Wintner, Lisa M. and Regnault, Antoine and {Ingelgård}, Anders and Coens, Corneel and Cessie, Saskia le and Holzner, Bernhard and Chang, Jane and Taphoorn, Martin and Cislo, Paul and Giesinger, Johannes M. and Cappelleri, Joseph C. and Pawar, Vivek and Seldam, Silene ten and Papadopoulos, Elektra J. and Calvert, Melanie J. and Joseph, Kenth L. and Bottomley, Andrew and Griebsch, Ingolf and Arraras, Juan I. and Astrup, Guro L. and Basch, Ethan and {Belan{\v{c}}i{\'{c}}}, Andrej and Brundage, Michael and Campbell, Alicyn and Rantell, Khadija Rerhou and Cocks, Kim and Cherny, Nathan and Eremenco, Sonya and Ferrer, Montse and Fiero, Mallorie H. and Gerlinger, Christoph and Goetghebeur, Els and Grouven, Ulrich and Lauer, Alexandra and Aiyegbusi, Olalekan Lee and Machingura, Abigirl and Mizusawa, Junki and Molenberghs, Geert and Petersen, Morten Aa and Reijneveld, Jaap C. and Ringash, Jolie and Rumpold, Gerhard and Rutherford, Claudia and Quinten, Chantal and Sail, Kavita and Sasseville, Maxime and Sauerbrei, Willi and Schiel, Anja and Smith, Ashley W. and Snyder, Claire and Velikova, Galina and Wang, Xin S. and Bjordal, Kristin and Pe, Madeline and Alexi, Solomon and Al-naesan, Imad and Black, Jennifer and Boele, Florien and Bogaerts, Kris and {Böhme}, Sarah and Chen, Ting-Yu and Chiavaro, Francesca and Claus, Vincent and Cleeland, Charlie and Coon, Cheryl and Rivera, Samantha Cruz and Arbanas, Gracia Dekanic and Desplanques, Gregoire and Emechebe, Nnadozie and Fitzgerald, Kristina and Garin, Olatz and Ge, Miaomiao and Gilbert, Alexandra and Groenvold, Mogens and Gross, Franziska and Horodniceanu, Erica and Johnson, Laura Lee and Joyner, Katie and Kamalakar, Rajesh and King, Madeleine and Kluetz, Paul and Kuljanic, Karin and Latino, Nicola and Liu, Limin and Maguri, Morine and Martinelli, Francesca and Mazerolle, Flora and Mitchell, Sandra and Morgan, Kate and Musoro, Jammbe and Ness, David and Nsiah, Irene and Oosting, Sjoukje and Peacock, Rosemary},
	year = {2025},
	month = {12},
	date = {2025-12-01},
	journal = {The Lancet Oncology},
	pages = {e683--e693},
	volume = {26},
	number = {12},
	doi = {10.1016/S1470-2045(25)00520-0}
}

@article{wen2018a,
	title = {Methods for handling longitudinal outcome processes truncated by dropout and death},
	author = {Wen, Lan and Terrera, Graciela Muniz and Seaman, Shaun R},
	year = {2018},
	month = {10},
	date = {2018-10-01},
	journal = {Biostatistics},
	pages = {407--425},
	volume = {19},
	number = {4},
	doi = {10.1093/biostatistics/kxx045}
}

@article{wen2018b,
	title = {Semi-Parametric Methods of Handling Missing Data in Mortal Cohorts Under Non-Ignorable Missingness},
	author = {Wen, Lan and Seaman, Shaun R.},
	year = {2018},
	month = {12},
	date = {2018-12-01},
	journal = {Biometrics},
	pages = {1427--1437},
	volume = {74},
	number = {4},
	doi = {10.1111/biom.12891}
}

@article{xiang2023,
	title = {The survival-incorporated median vs the median in the survivors or in the always-survivors: What are we measuring? and Why?},
	author = {Xiang, Qingyan and Bosch, Ronald J. and Lok, Judith J.},
	year = {2023},
	month = {12},
	date = {2023-12-20},
	journal = {Statistics in medicine},
	pages = {5479--5490},
	volume = {42},
	number = {29},
	doi = {10.1002/sim.9922}
}

@article{pham2025,
	title = {Estimating treatment effects in trials with outcome data truncated by death: A case study on aligning estimators with estimands},
	author = {Pham, Tra My and Kahan, Brennan C and Lopes, Andre and Rashid, Memuna and Hoskin, Peter J and White, Ian R},
	year = {2025},
	month = {12},
	date = {2025-12-01},
	journal = {Clinical Trials},
	pages = {676--686},
	volume = {22},
	number = {6},
	doi = {10.1177/17407745251360645},
	langid = {EN}
}

@article{dufouil2004,
	title = {Analysis of longitudinal studies with death and drop-out: a case study},
	author = {Dufouil, Carole and Brayne, Carol and Clayton, David},
	year = {2004},
	date = {2004},
	journal = {Statistics in Medicine},
	pages = {2215--2226},
	volume = {23},
	number = {14},
	doi = {10.1002/sim.1821},
	url = {https://onlinelibrary.wiley.com/doi/abs/10.1002/sim.1821},
	note = {{\_}eprint: https://onlinelibrary.wiley.com/doi/pdf/10.1002/sim.1821},
	langid = {en}
}

@misc{vanderpijpen,
	title={Improved prediction of extreme random effects in joint models: WRaPs}, 
	author={Vanderpijpen, Eline and Goetghebeur, Els},
	year={2026},
	eprint={2606.18843},
	archivePrefix={arXiv},
	primaryClass={stat.ME},
	url={https://arxiv.org/abs/2606.18843}, 
}

\appendix

\section*{Acknowledgements}

	We thank the wider SISAQOL-IMI consortium for many valuable discussions and the infrastructure that enabled this research. 

We also thank the STRATOS initiative for their valuable methodological input during this study, and their members' involvement in SISAQOL-IMI Work Package 3.

This publication reflects the views of the individual authors and should not be construed to represent official views or policies of the European Medicines Agency (EMA), the US Food and Drug Administration (FDA), US National Cancer Institute (NCI), Medicines and Healthcare products Regulatory Agency (MHRA), Institute for Quality and Efficiency in Health Care (IQWiG), Health Canada, the Norwegian Medicines Agency (NOMA), the American Society of Clinical Oncology (ASCO) or the European Society for Medical Oncology (ESMO) or any other institution, organization, or entity. This publication reflects the author's view and neither IMI nor the EU, EFPIA are responsible for any use that may be made of the information contained therein. SISAQOL-IMI Work Package 3-members involved in the paper: \\

Dries Reynders$^1$, Doranne Thomassen$^2$, Satrajit Roychoudhury$^3$, Cecilie Delphin Amdal$^{4,5}$, Jammbe Z. Musoro$^6$, Willi Sauerbrei$^7$, Saskia le Cessie$^{1,2,8}$, Els Goetghebeur$^1$ and Ethan Basch$^{9}$, Melanie Calvert$^{10,11}$, Joseph C. Cappelleri$^2$, Samantha Cruz Rivera$^{10}$, Olalekan Lee Aiyegbusi$^{10,11}$, Geert Molenberghs$^{12}$, Khadija Rerhou Rantell$^{13}$, Alexander Russell-Smith$^{14}$, Anja Schiel$^{15}$, Ragnild Sorum Falk$^{3}$\\

$^9$ American Society for Clinical Oncology, Alexandria, VA, USA\\
$^{10}$ University of Birmingham, Birmingham, UK\\
$^{11}$ NIHR Birmingham Biomedical Research Centre, Birmingham, UK\\
$^{12}$ Katholieke Universiteit Leuven, Leuven, Belgium\\
$^{13}$ Medicines and Healthcare products Regulatory Agency, London, UK\\
$^{14}$ Pfizer Ltd., Sandwich, UK\\
$^{15}$ Norwegian Medicines Agency, Oslo, Norway

\section{While-Alive - Illustration}\label{secA1}

\begin{figure}
	\centering{
		\includegraphics[width = 0.9 \textwidth]{%figattritcens5-1.pdf
			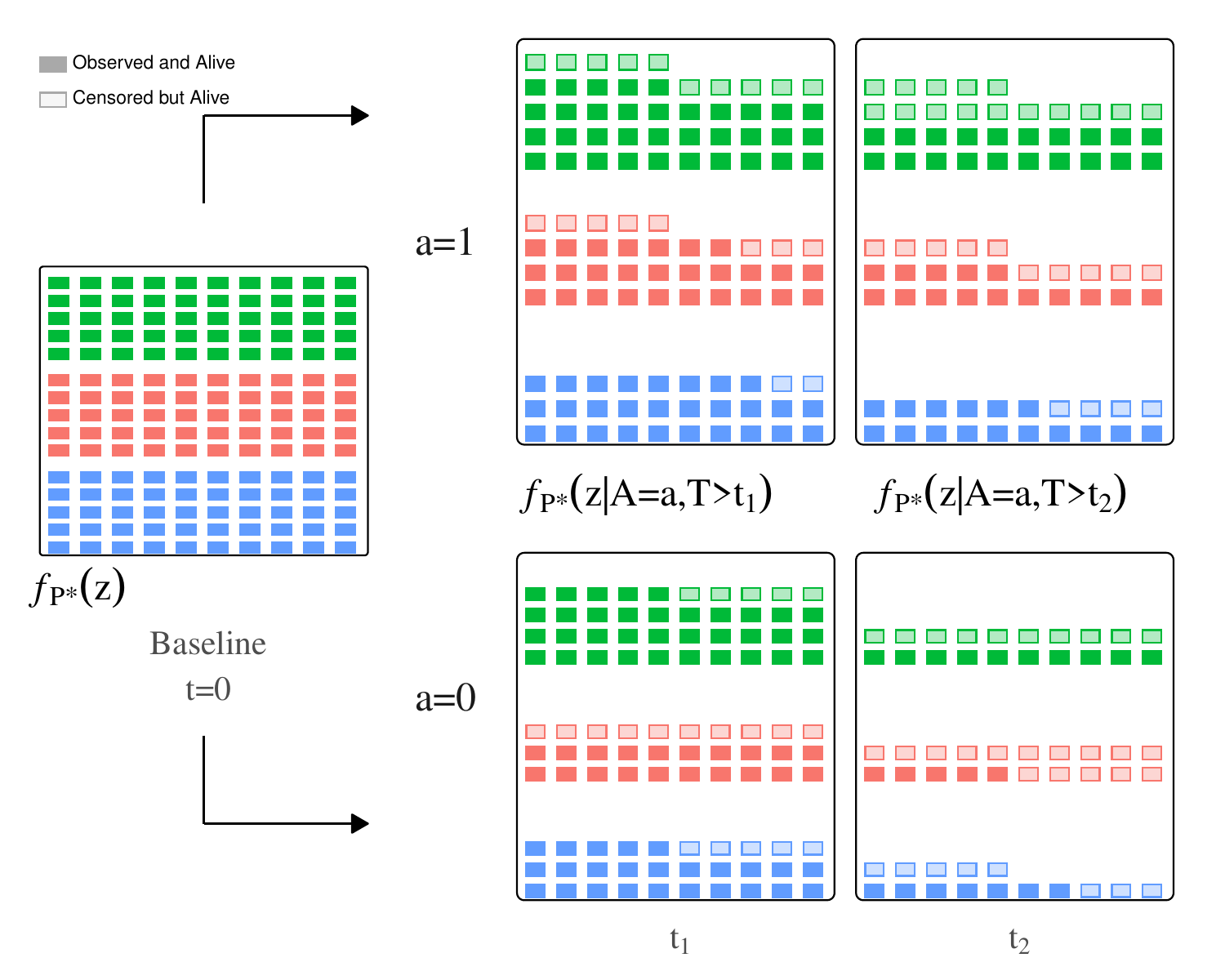}
	}
	\caption{\footnotesize All blocks represent patients still alive, but a shaded block means they are censored and hence in reality we do not know whether they are still alive or not. Colors  represent different values for their baseline-confounders $\mathbf{Z}$. We can imagine that green patients tend to have better longitudinal outcomes than red patients who in turn tend to do better than blue. Observed results under both treatments $A=0/1$, are transported to the baseline \(\mathbf{Z}\)-distribution of target population $\cal{P}^*$. The different number of blocks of each color at times $t_1$ and $t_2$ between the treatments represents differential survival ($P(T>t^*|A=a, \mathbf{Z})$) which leads to divergent $\mathbf{Z}$-distributions among those alive at $t^*$ ($f_{\cal{P}^*}(z|A=a, T>t^*)$) under treatments $a=0$ and $a=1$. The result of our while-alive estimand does not only depend on (the trend of) the outcome within each colour-profile, but also on the relative sizes of the green, red and blue group at each time point. As can be seen from the shaded blocks, due to censoring, the observed \(\mathbf{Z}\)-distribution under those alive at $t^*$, \(f_{\cal{P}^*}(z|X>t^*)\), may differ from the actual distribution among survivors \(f_{\cal{P}^*}(z|T>t^*)\), while it is the latter we are interested in.}
	\label{fig-scheme1b}
\end{figure}

Figure \ref{fig-scheme1b} provides a graphical representation of selection through survival impacts the while-alive estimand and how censoring needs to be addressed to properly estimate it.

\section{Simulation}

\label{sec-appsim}

R-code is available at https://github.com/DRUgent/WhileAlive. Appendix Table \ref{tbl-bias} presents the empirical bias and SD for all evaluated methods.

The data generating mechanism for the results presented in the paper is:

\begin{itemize}
	\item
	\(Z_i\sim \mathcal{U}(-1; 1)\)
	\item
	\(\mathrm{logit}\left(P(A_i = 1|Z_i)\right) = 0 +1 \cdot Z_i\)
	\item
	$Y_{ij} = Y_i(t_j) = 0 \cdot A_i+1 \cdot Z_i + 0.2\cdot \mathrm{t_j}+0.2 \cdot A_i \cdot Z_i+0.2 \cdot A_i \cdot t_j + u_i + \epsilon_{ij}$ with $t_j=j=0,\ldots, 10$ and a person-specific intercept $u_i \sim \mathcal{N}(0,1)$ and error $\epsilon_{ij} \sim \mathcal{N}(0,1)$ independent from each other.
	\item
	\(\lambda_T(t,A_i,Z_i, \mathbf{\overline{Y}}_i(t)) =\lambda_T(t,A_i,Z_i, Y_i(\lfloor t \rfloor))\) \\ \(= 0.0625\cdot e^{-0.36 \cdot A_i  -0.36 \cdot Z_i -0.22 \cdot Y_i(\lfloor t \rfloor) + 0.18 \cdot A_i\cdot Z_i + 0.26 \cdot A_i \cdot Y_i(\lfloor t \rfloor)}\) where $Y(\lfloor t \rfloor)=Y(t_m)$ with $t_m$ the most recent $t_j$ preceding $t$
	\item
	\(\lambda_C(t,A_i,Z_i, \mathbf{\overline{Y}}_i(t)) = \lambda_C(t,A_i,Z_i, Y_i(\lfloor t \rfloor))\) \\ \(= 0.05\cdot e^{0 \cdot A_i  -0.11 \cdot Z_i  -0.69 \cdot Y_i(\lfloor t \rfloor)  -0.11 \cdot A_i \cdot Z_i + 0.92 \cdot A_i \cdot Y_i(\lfloor t \rfloor)}\)
\end{itemize}
In the analyses, data of patients $i$ at time $t_j$ were only used if $t_j <$min$(T_i, C_i, 10.1)$.

The implemented methods are:

\begin{itemize}
	
	\item
	\textbf{Naive Time Point Specific}
	
	\begin{itemize}	
		\item
		\textbf{As Observed}: Difference in observed average \(Y\) between the two treatment groups as estimated with IEE.
		\item
		\textbf{IPCW only}: Difference in IPC weighted average \(Y\) between the two treatment groups as estimated with IEE, with weights derived from arm specific Cox-models for censoring with $Y(t)$ as time-varying covariate.
		\item
		\textbf{IPTW only}: Difference in IPT weighted average \(Y\) between the two treatment groups as estimated by IEE
	\end{itemize}
	\item
	\textbf{Adapted Methods - Baseline Censoring only}
	
	\begin{itemize}
		\item
		\textbf{IEE-IPTCW (no TV Cens)}: IPTC weighted IEE. One Cox regression model for censoring for all time points (separate per treatment group) - time considered as categorical variable in the IEE model. (Section 3.3). Censoring model with baseline-covariates only
		\item
		\textbf{IEE-Regression Standardization (no TV Cens)}: One Cox regression model for death for all time points (separate per treatment group). IEE for \(E(Y(t)|Z, T>t)\) for both arms combined (Section 3.3), with treatment-time and treatment-$Z$ interaction and categorical time. No censoring weights included, as censoring is assumed to depend only on baseline-covariates already in the Y- and Survival-models.
	\end{itemize}
	\item
	\textbf{Adapted Longitudinal - Time varying censoring}
	
	\begin{itemize}
		\item
		\textbf{IEE-IPTCW}: IPTC weighted IEE. One Cox regression model for censoring with $Y(t$ as time-varying covariate (separate per treatment group) - time considered as categorical variable in the IEE model. (Section 3.3)
		\item
		\textbf{IEE-Regression Standardization}: One Cox regression model for death for all time points (separate per treatment group). IEE for \(E(Y(t)|Z, T>t)\) for both arms combined (Section 3.3), with treatment-time and treatment-$Z$ interaction and categorical time. Both models are IPCWeighted with the same weights as the IPTCW-model.
	\end{itemize}
\end{itemize}

\begin{table}[h]
	\centering\begingroup\fontsize{7}{9}\selectfont
	\caption{\label{tbl-bias}Bias and SD for the different estimators in the simulation study. Coverage if for percentile bootstrap- based confidence intervals}
	\begin{tabular}{lllllllllllll}
		\toprule
		Label & M. & T 0 & T 1 & T 2 & T 3 & T 4 & T 5 & T 6 & T 7 & T 8 & T 9 & T 10\\
		\midrule
		As Observed & Bias & 0.32 & 0.25 & 0.16 & 0.08 & 0.02 & 0.01 & -0.05 & -0.09 & -0.15 & -0.19 & -0.22\\
		& SD & 0.11 & 0.12 & 0.12 & 0.13 & 0.14 & 0.15 & 0.15 & 0.16 & 0.17 & 0.17 & 0.20\\
		& Coverage & 0.15 & 0.40 & 0.72 & 0.90 & 0.94 & 0.93 & 0.94 & 0.91 & 0.83 & 0.81 & 0.74\\
		IPCW Only & Bias & 0.32 & 0.34 & 0.33 & 0.30 & 0.29 & 0.31 & 0.30 & 0.29 & 0.27 & 0.26 & 0.27\\
		& SD & 0.11 & 0.12 & 0.12 & 0.14 & 0.15 & 0.15 & 0.16 & 0.17 & 0.18 & 0.18 & 0.22\\
		\addlinespace
		& Coverage & 0.15 & 0.17 & 0.25 & 0.38 & 0.47 & 0.43 & 0.54 & 0.59 & 0.69 & 0.74 & 0.72\\
		IPTW only & Bias & 0.00 & -0.05 & -0.12 & -0.19 & -0.24 & -0.25 & -0.30 & -0.33 & -0.39 & -0.42 & -0.44\\
		& SD & 0.10 & 0.12 & 0.12 & 0.14 & 0.14 & 0.15 & 0.15 & 0.17 & 0.18 & 0.18 & 0.21\\
		& Coverage & 0.96 & 0.93 & 0.84 & 0.68 & 0.58 & 0.58 & 0.50 & 0.47 & 0.38 & 0.35 & 0.38\\
		IEE-IPTCW (no TV Cens) & Bias & 0.00 & -0.03 & -0.09 & -0.15 & -0.20 & -0.20 & -0.24 & -0.27 & -0.33 & -0.35 & -0.37\\
		\addlinespace
		& SD & 0.10 & 0.11 & 0.12 & 0.13 & 0.14 & 0.15 & 0.15 & 0.16 & 0.17 & 0.17 & 0.20\\
		& Coverage & 0.96 & 0.94 & 0.87 & 0.76 & 0.68 & 0.69 & 0.63 & 0.61 & 0.48 & 0.46 & 0.47\\
		IEE-RegStand (no TV Cens) & Bias & 0.04 & -0.01 & -0.08 & -0.15 & -0.20 & -0.20 & -0.25 & -0.29 & -0.34 & -0.37 & -0.39\\
		& SD & 0.10 & 0.11 & 0.11 & 0.13 & 0.13 & 0.14 & 0.14 & 0.15 & 0.16 & 0.16 & 0.19\\
		& Coverage & 0.94 & 0.94 & 0.89 & 0.74 & 0.65 & 0.67 & 0.59 & 0.53 & 0.43 & 0.40 & 0.42\\
		\addlinespace
		IEE-IPTCW & Bias & 0.00 & 0.03 & 0.02 & 0.00 & -0.01 & 0.02 & 0.01 & 0.01 & -0.02 & -0.02 & -0.01\\
		& SD & 0.10 & 0.11 & 0.12 & 0.14 & 0.15 & 0.15 & 0.15 & 0.16 & 0.17 & 0.18 & 0.21\\
		& Coverage & 0.96 & 0.94 & 0.95 & 0.94 & 0.94 & 0.93 & 0.95 & 0.95 & 0.93 & 0.94 & 0.93\\
		IEE-RegStand & Bias & 0.01 & 0.04 & 0.03 & 0.00 & -0.01 & 0.02 & 0.01 & 0.00 & -0.02 & -0.03 & -0.02\\
		& SD & 0.10 & 0.11 & 0.12 & 0.13 & 0.14 & 0.15 & 0.15 & 0.16 & 0.17 & 0.18 & 0.21\\
		\addlinespace
		& Coverage & 0.96 & 0.93 & 0.95 & 0.95 & 0.94 & 0.94 & 0.95 & 0.96 & 0.93 & 0.93 & 0.92\\
		\bottomrule
	\end{tabular}
	
	\endgroup
\end{table}

\clearpage

\section{Case Study}

\begin{figure}
	\centering{
		\includegraphics[width = \textwidth, height=0.8\textheight]{%figattritcens5-1.pdf
			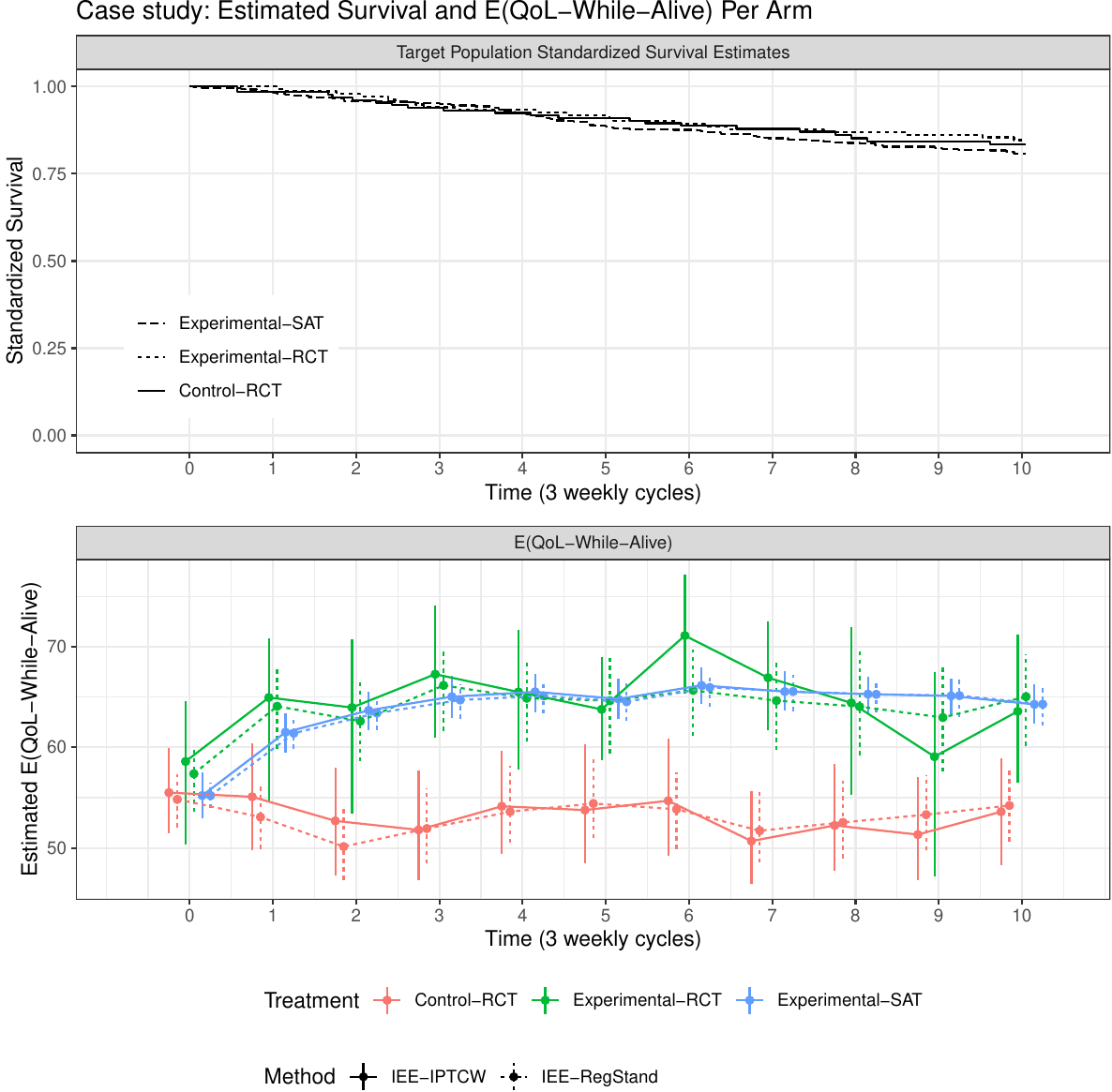}
	}
	\caption{\footnotesize Case Study - estimated target population-survival and average while-alive HRQoL for the SAT and the two RCT-arms. Confidence intervals are bootstrap based.}
	\label{fig-perarm}
\end{figure}

Figure \ref{fig-perarm} shows the estimated target population-survival and average while-alive HRQoL for the SAT and the two RCT-arms.

Appendix Tables \ref{tbl-demcomp} and \ref{tbl-demred} provide an overview of the considered baseline distributions in the complete sample and after reduction for non-positivity.

For the analyses, data is prepared in the tstart-tstop-format required by the R-survival package to handle time-varying covariates. This allows the censoring model to be fitted regardless of whether time-varying covariates are included or not, from which then estimated probabilities of censoring are derived and the corresponding IPC-weights. For the single-time point analyses, only data up till that time point is included in the censoring-model. As for early time points there are few censoring-events, this sometimes results in issues where the survival package manages to fit a cox-model but does not provide derived censoring probabilities. As this is most likely to happen in a model with time varying covariates, first the covariate 'progression' is dropped from the model in an attempt to solve the issue. If that doesn't work, IPC-weights are set to 1 (this will only happen with very few censoring events). For the longitudinal (IEE-based) analyses, the IPC weights are estimated from one analysis of the complete follow up. Hence, no such issues occurred in that setting.

The results of the CoxPH-censoring models with time varying covariates over the 30-week follow up, are presented below.

IPT-weights are estimated using the weightit-function of the WeightIt-package. Weights are trimmed at the 99th percentile. 

Confidence intervals are based on 200 bootstraps.

\begin{table}[htbp]
	\centering
	\caption{\label{tbl-demcomp}Baseline demographics - Complete sample}
	\begin{tabular}{lccc}
		\toprule
		& \multicolumn{3}{c}{Complete sample} \\
		\cmidrule(l{3pt}r{3pt}){2-4}
		\textbf{Characteristic} 
		& \parbox{3cm}{\centering \textbf{Experimental - SAT}\\ N = 879}
		& \parbox{3cm}{\centering \textbf{Control - RCT}\\ N = 158}
		& \parbox{3cm}{\centering \textbf{Experimental - RCT}\\ N = 162} \\
		\midrule
		Sex & & & \\
		\hspace{1em}Female & 503 (57\%) & 89 (56\%) & 88 (54\%)\\
		\hspace{1em}Male   & 376 (43\%) & 69 (44\%) & 74 (46\%)\\
		Age & 53 (44, 62) & 49 (39, 59) & 51 (41, 60)\\
		Baseline ECOG Score & & & \\
		\addlinespace
		\hspace{1em}0 & 249 (28\%) & 63 (40\%) & 72 (44\%)\\
		\hspace{1em}1 & 480 (55\%) & 84 (53\%) & 75 (46\%)\\
		\hspace{1em}2 & 119 (14\%) & 11 (7.0\%) & 15 (9.3\%)\\
		\hspace{1em}3 & 28 (3.2\%) & 0 (0\%) & 0 (0\%)\\
		\hspace{1em}Unknown & 3 & 0 & 0\\
		\addlinespace
		Number of Prior Drug Therapy Regimens & & & \\
		\hspace{1em}1 & 181 (21\%) & 128 (81\%) & 142 (88\%)\\
		\hspace{1em}2 & 323 (37\%) & 29 (18\%) & 17 (10\%)\\
		\hspace{1em}3 & 189 (22\%) & 1 (0.6\%) & 3 (1.9\%)\\
		\hspace{1em}$\geq$4 & 183 (21\%) & 0 (0\%) & 0 (0\%)\\
		\addlinespace
		\hspace{1em}Unknown & 3 & 0 & 0\\
		baseqol & 50 (33, 75) & 58 (42, 75) & 58 (42, 75)\\
		\hspace{1em}Unknown & 41 & 10 & 7\\
		\bottomrule
		\multicolumn{4}{l}{\textsuperscript{1} n (\%); Median (IQR)}\\
	\end{tabular}
\end{table}

\begin{table}[htbp]
	\centering
	\caption{\label{tbl-demred}Baseline demographics - Reduced for positivity}
	\begin{tabular}{lccc}
		\toprule
		& \multicolumn{3}{c}{Reduced for Positivity} \\
		\cmidrule(l{3pt}r{3pt}){2-4}
		\textbf{Characteristic} 
		& \parbox{3cm}{\centering \textbf{Experimental - SAT}\\ N = 467}
		& \parbox{3cm}{\centering \textbf{Control - RCT}\\ N = 147}
		& \parbox{3cm}{\centering \textbf{Experimental - RCT}\\ N = 152} \\
		\midrule
		Sex & & & \\
		\hspace{1em}Female & 272 (58\%) & 82 (56\%) & 82 (54\%)\\
		\hspace{1em}Male   & 195 (42\%) & 65 (44\%) & 70 (46\%)\\
		Age & 53 (42, 63) & 49 (41, 59) & 51 (41, 59)\\
		Baseline ECOG Score & & & \\
		\addlinespace
		\hspace{1em}0 & 154 (33\%) & 57 (39\%) & 68 (45\%)\\
		\hspace{1em}1 & 248 (53\%) & 80 (54\%) & 70 (46\%)\\
		\hspace{1em}2 & 65 (14\%) & 10 (6.8\%) & 14 (9.2\%)\\
		\addlinespace
		Number of Prior Drug Therapy Regimens & & & \\
		\hspace{1em}1 & 169 (36\%) & 121 (82\%) & 135 (89\%)\\
		\hspace{1em}2 & 298 (64\%) & 26 (18\%) & 17 (11\%)\\
		baseqol & 58 (33, 75) & 58 (42, 75) & 58 (42, 75)\\
		\bottomrule
		\multicolumn{4}{l}{\textsuperscript{1} n (\%); Median (IQR)}\\
	\end{tabular}
\end{table}

\clearpage

\subsection*{Outcome Censoring Models}\label{sec-censoringmodels}

The tables below present the outcome of the CoxPH-models with time varying covariates for censoring in the primary analysis and the negative control analysis. \textit{QoL} is the time varying observed Health Related Quality of Life at the start of each 3-weeks cycle. \textit{Prog} is a time varying indicator for progression.

Note that \textit{n=} in the description below the tables refers to the number of rows in the dataset and not the number of patients, given the tstart-tstop format.

The censoring in both RCT-arms is around 20\%, the censoring in the SAT is considerably lower.

\subsubsection*{Primary Analysis}

\begin{table}[htbp]
	\centering
	\caption{Fitting Proportional Hazards Regression Model for Censoring: SAT Experimental}
	\begin{tabular}{lccccc}
		\toprule
		& coef & exp(coef) & se(coef) & z & p \\
		\midrule
		\textbf{sexMALE} & -0.7271 & 0.4833 & 0.5901 & -1.232 & 0.2179 \\
		\textbf{Age} & 0.01839 & 1.019 & 0.0182 & 1.01 & 0.3123 \\
		\textbf{ECOGBfactor1} & -0.3376 & 0.7135 & 0.604 & -0.5589 & 0.5762 \\
		\textbf{ECOGBfactor2} & 0.7536 & 2.125 & 0.7105 & 1.061 & 0.2889 \\
		\textbf{NPSTREGSfactor2} & -0.6481 & 0.523 & 0.5126 & -1.264 & 0.2061 \\
		\textbf{QoL} & -0.01885 & 0.9813 & 0.01302 & -1.447 & 0.1478 \\
		\textbf{Prog} & 1.429 & 4.174 & 0.5698 & 2.508 & 0.01215 \\
		\bottomrule
	\end{tabular}
	Likelihood ratio test=17.26 on 7 df, p=0.0158042 n= 5099, number of events= 16
\end{table}

\begin{table}[htbp]
	\centering
	\caption{Fitting Proportional Hazards Regression Model for Censoring: RCT Control}
	\begin{tabular}{lccccc}
		\toprule
		& coef & exp(coef) & se(coef) & z & p \\
		\midrule
		\textbf{sexMALE} & 0.7306 & 2.076 & 0.3891 & 1.878 & 0.0604 \\
		\textbf{Age} & 0.0213 & 1.022 & 0.01408 & 1.512 & 0.1305 \\
		\textbf{ECOGBfactor1} & -0.1611 & 0.8512 & 0.3785 & -0.4256 & 0.6704 \\
		\textbf{ECOGBfactor2} & -0.2442 & 0.7833 & 1.067 & -0.229 & 0.8189 \\
		\textbf{NPSTREGSfactor2} & -0.6433 & 0.5255 & 0.5458 & -1.179 & 0.2385 \\
		\textbf{QoL} & -0.01266 & 0.9874 & 0.009441 & -1.34 & 0.1801 \\
		\textbf{Prog} & -0.4276 & 0.6521 & 0.4015 & -1.065 & 0.2868 \\
		\bottomrule
	\end{tabular}
	Likelihood ratio test=7.89 on 7 df, p=0.3426893 n= 1522, number of events= 30
\end{table}

\clearpage

\subsubsection*{Negative Control Analysis}

\begin{table}[htbp]
	\centering
	\caption{Fitting Proportional Hazards Regression Model for Censoring: SAT Experimental}
	\begin{tabular}{lccccc}
		\toprule
		& coef & exp(coef) & se(coef) & z & p \\
		\midrule
		\textbf{sexMALE} & -0.7271 & 0.4833 & 0.5901 & -1.232 & 0.2179 \\
		\textbf{Age} & 0.01839 & 1.019 & 0.0182 & 1.01 & 0.3123 \\
		\textbf{ECOGBfactor1} & -0.3376 & 0.7135 & 0.604 & -0.5589 & 0.5762 \\
		\textbf{ECOGBfactor2} & 0.7536 & 2.125 & 0.7105 & 1.061 & 0.2889 \\
		\textbf{NPSTREGSfactor2} & -0.6481 & 0.523 & 0.5126 & -1.264 & 0.2061 \\
		\textbf{QoL} & -0.01885 & 0.9813 & 0.01302 & -1.447 & 0.1478 \\
		\textbf{Prog} & 1.429 & 4.174 & 0.5698 & 2.508 & 0.01215 \\
		\bottomrule
	\end{tabular}
	Likelihood ratio test=17.26 on 7 df, p=0.0158042 n= 5099, number of events= 16
\end{table}

\begin{table}[htbp]
	\centering
	\caption{Fitting Proportional Hazards Regression Model for Censoring: RCT Experimental}
	\begin{tabular}{lccccc}
		\toprule
		& coef & exp(coef) & se(coef) & z & p \\
		\midrule
		\textbf{sexMALE} & 0.03874 & 1.04 & 0.3768 & 0.1028 & 0.9181 \\
		\textbf{Age} & 0.03844 & 1.039 & 0.01448 & 2.655 & 0.007942 \\
		\textbf{ECOGBfactor1} & -0.3257 & 0.722 & 0.3985 & -0.8175 & 0.4137 \\
		\textbf{ECOGBfactor2} & 0.1583 & 1.171 & 0.702 & 0.2255 & 0.8216 \\
		\textbf{NPSTREGSfactor2} & -0.1102 & 0.8956 & 0.562 & -0.1962 & 0.8445 \\
		\textbf{QoL} & -0.005568 & 0.9944 & 0.009365 & -0.5945 & 0.5521 \\
		\textbf{Prog} & 0.2844 & 1.329 & 0.4761 & 0.5974 & 0.5502 \\
		\bottomrule
	\end{tabular}
	Likelihood ratio test=11.86 on 7 df, p=0.105344 n= 1542, number of events= 34
\end{table}

\end{document}